\documentclass[twocolumn]{aastex631}
\usepackage[flushleft]{threeparttable}
\usepackage{lmodern}

\begin{document}

\title{Medium Resolution 0.97-5.3~$\mu$m spectra of Very Young Benchmark Brown Dwarfs with NIRSpec onboard the James Webb Space Telescope}

\author[0000-0003-0192-6887]{Elena Manjavacas}
\affiliation{AURA for the European Space Agency (ESA), ESA Office, Space Telescope Science Institute, 3700 San Martin Drive, Baltimore, MD, 21218 USA}
\affiliation{Department of Physics and Astronomy, Johns Hopkins University, Baltimore, MD 21218, USA}
\correspondingauthor{Elena Manjavacas}
\email{emanjavacas@stsci.edu}

\author[0000-0001-6172-3403]{Pascal Tremblin}
\affiliation{Universit\'e Paris-Saclay, UVSQ, CNRS, CEA, Maison de la Simulation, F-91191, Gif-sur-Yvette, France}

\author{Stephan Birkmann}
\affiliation{European Space Agency, European Space Astronomy Centre, Camino Bajo del Castillo s/n, E-28692 Villanueva de la Ca\~nada, Madrid, Spain}

\author{Jeff Valenti}
\affiliation{Space Telescope Science Institute, 3700 San Martin Drive, Baltimore, MD 21218, USA}

\author[0000-0003-2896-4138]{Catarina Alves de Oliveira}
\affiliation{European Space Agency, European Space Astronomy Centre, Camino Bajo del Castillo s/n, E-28692 Villanueva de la Ca\~nada, Madrid, Spain}

\author[0000-0002-6881-0574]{Tracy L. Beck}
\affiliation{Space Telescope Science Institute, 3700 San Martin Drive, Baltimore, MD 21218, USA}

\author[0000-0002-9262-7155]{G. Giardino} 
\affiliation{ATG Europe for the European Space Agency, ESTEC, Keplerlaan 1, 2200 AG Noordwijk, Netherlands}

\author[0000-0002-4034-0080]{N. L\"utzgendorf} 
\affiliation{European Space Agency, ESTEC, Keplerlaan 1, 2200 AG Noordwijk, Netherlands}

\author[0000-0003-2662-6821]{B. J. Rauscher} 
\affiliation{NASA Goddard Space Flight Center, Observational Cosmology Laboratory, Greenbelt, USA}

\author{M. Sirianni} 
\affiliation{European Space Agency (ESA), ESA Office, Space Telescope Science Institute, 3700 San Martin Drive, Baltimore, MD 21218, USA}



\begin{abstract}

Spectra of young benchmark brown dwarfs with well-known ages are vital to characterize other brown dwarfs, for which ages are in general not known. These spectra are also crucial to test atmospheric models which have the potential to provide detailed information about the atmospheres of these objects. However, to {optimally} test atmospheric {models}, medium-resolution, long-wavelength coverage spectra with well-understood uncertainties are {ideal}, such as the spectra provided by the NIRSpec instrument onboard the \textit{James Webb Space Telescope}. In this paper, we present the medium-resolution JWST/NIRSpec spectra of two young brown dwarfs, TWA~28 {(M9.0)} and TWA~27A {(M9.0)}, and one planetary-mass object, TWA~27B {(L6.0)}, members of the TW Hydrae Association ($\sim$10~Myr). We show the richness of the atomic lines and molecular bands present in the spectra. All objects show signs of a circumstellar disk, via near-infrared excess and/or via emission lines. We matched a set of cloudless atmospheric spectra ({\tt ATMO}), and cloudy atmospheric spectra ({\tt BT-Settl}) to our NIRSpec spectra, and analyzed which wavelength ranges and spectral features both models reproduce best. Both models derive consistent parameters for the three sources, and predict the existence of $\mathrm{CH_{4}}$ at 3.35 $\mu$m in TWA~27B. Nonetheless, in contrast to other slightly older objects with similar spectral type, like PSO~318.5--22 and VHS~1256b, this feature is not present in the spectrum of TWA~27B. The lack of the $\mathrm{CH_{4}}$ feature might suggest that the L/T transition of very young dwarfs starts at later spectral types than for older brown dwarfs.

\end{abstract}



\section{Introduction} \label{sec:intro}

{Brown dwarfs are canonically defined as  substellar objects {intermediate} between low-mass stars and planets ($\mathrm{13~M_{Jup}<M<75~M_{Jup}}$)}, {never reaching high enough core temperature to fuse hydrogen}. Thus, since they are born, brown dwarfs cool down with time \citep{Burrows, Baraffe}. Due to their evolution, given any brown dwarf, it is challenging in general to determine its age or its mass (age-mass degeneracy), which turns the {accurate} physical characterization of brown dwarfs into a {challenge}, {since most brown dwarfs are free-floating, and therefore their ages and/or masses are highly unconstrained}. For this reason, spectra of brown dwarfs with well-determined ages, i.e.  members {of} open clusters, young associations, young moving groups, or companions {to higher-mass objects}, are of key importance for the characterization of brown dwarfs as a whole. 

Dozens of spectra of young benchmark brown dwarfs, younger than $\sim$200~Myr, and with a reliable estimation of their age {(i.e. companions to stars, bona fide members of young moving groups, members of open cluster associations)}, were published in the literature since the discovery of brown dwarfs (e.g. \citealt{Allers2013, Bonnefoy2014a, Manjavacas2014, Faherty2016, Manjavacas2020, Miles2023}). Nevertheless, most of those spectra are limited to the optical and the near-infrared in the best cases, and the majority have been obtained using ground-based facilities, which limits the precision of their calibrations, due to the fast-varying nature of telluric absorptions of the Earth's atmosphere. Although tellurics can be corrected up to some extent, their removal is usually not perfect. In addition, direct comparison between brown dwarf spectra of the same objects obtained with ground-based facilities and the \textit{Hubble Space Telescope} show differences in colors and in the shape of the water band at 1.4~$\mu$m \citep{Manjavacas2019}. 

Although ground-based brown dwarf spectra remain extremely useful, they are not ideal for atmospheric model testing, {since the majority just cover near-infrared wavelengths or near-infrared and optical wavelengths in the best case, providing a limited coverage of their spectral energy distribution (SED). In addition, there are few ground-based brown dwarf spectral libraries that cover the mid-infrared. Having a {wide} wavelength coverage is particularly important for deriving accurate bolometric luminosities, and for testing synthetic spectra derived using atmospheric models. In particular for testing retrieval models, having, in addition, reliable flux uncertainties is crucial}, since their results rely on the uncertainties of the spectra being accurate to enable a probabilistic assessment of their fitted parameters and their uncertainties. Thus, medium-resolution, accurately-calibrated, long-wavelength coverage spectra, like those provided by the recently commissioned near- and mid-infrared instruments onboard the \textit{James Webb Space Telescope} (JWST) are needed to properly test atmospheric models.

{Finally, even though addressing this question is beyond the scope of this work, atmospheric models allow us to provide further constraints on the formation mechanism for brown dwarfs and planets via the measurement of the C/O ratios. A brown dwarf atmosphere could have a supersolar  C/O ratio if it formed via core or pebble accretion, but will have a solar C/O ratio if formed via gravitational instability \citep{Orberg2011}. Carbon- and oxygen-bearing molecules ($\mathrm{H_{2}O}$, CO, $\mathrm{CH_{4}}$, and $\mathrm{CO_{2}}$) are the primary absorbers of the emitted spectra of brown dwarfs, and  the main carbon and oxygen-bearing species. For an accurate determination of C/O ratios long wavelength coverage spectra that allow us to measure the abundances of most of the carbon- and oxygen-bearing molecules in brown dwarf spectra are also crucial.}


{In this paper, we present the medium-resolution 0.97-5.3~$\mu$m NIRSpec/JWST spectra of two benchmark young M9.0 brown dwarfs members of the TW Hydrae association ($\sim$10$\pm$2~Myr, \citealt{Mamajek2005, Barrado2006, Luhman2023b}), TWA~28 and TWA~27A, and its planetary-mass companion, TWA~27B, which are the higher signal-to-noise, {broader} wavelength coverage, and higher resolution spectra of young benchmark brown dwarfs up date. We present a comparison to cloudy and cloudless atmospheric models. In Section \ref{sec:targets} we describe the main physical characteristics of the three brown dwarfs presented in this paper as presented in the literature. In Section \ref{sec:observations}, we describe the JWST/NIRSpec observations carried out for our targets. In Section \ref{sec:data_reduction} we describe the data reduction. In Section \ref{sec:data_analysis} we identify the atomic lines and molecular bands present in the spectra of TWA~28, TWA~27A and TWA~27B. In Section \ref{sec:atmospheric_models} we compare our spectra with cloudy and cloudless models, and derive their effective temperatures and surface gravities. In Section \ref{sec:discussion} we discuss our results. Finally, in Section \ref{sec:conclusions} we summarize our conclusions.}

\section{{Targets}}\label{sec:targets}

Both TWA~28 and TWA~27A were classified as M9.0 dwarfs by \cite{Venuti2019} using 0.30-2.50~$\mu$m X-shooter spectra (resolution R = 9700-11600), reporting very similar atmospheric and physical parameters for both objects. {\cite{Venuti2019} obtained X-Shooter spectra of eleven TWA members with infrared excess with spectral types between M0 and M9, and masses between 0.58~$M_{\odot}$ and 0.02~$M_{\odot}$. They derived the individual extinction, stellar parameters, and accretion parameters for each object simultaneously, and they measured the luminosity of Balmer lines and forbidden emission lines to probe the physics of the star-disk interaction environment.}

\cite{Venuti2019} derived an effective temperature of 2660~K, and a \textit{log~g} = 4.1, for TWA~28, with an estimated mass of 0.020$M_\odot$ (20.92~$M_{Jup}$), and a radius of 0.29~$R_{\odot}$ (2.9$R_{Jup}$), with a measured luminosity of \textit{log~L} = -2.47~$L_{\odot}$. Signs of a circumstellar disk were found for TWA~28 due to a near-infrared excess observed in its X-shooter spectrum, and emission lines {were} found mostly in UV and optical wavelengths. \cite{Venuti2019} measured an accretion rate estimate of $log(\dot{M}_{acc})$~=~-12.24 $M_{\odot}\,~yr^{-1}$.

For TWA~27A, \cite{Venuti2019} estimated an effective temperature of 2640~K, and a \textit{log~g} = 3.75, sightly smaller than for TWA~28. \cite{Venuti2019} estimated the mass of TWA~27A in 0.019~$M_\odot$ (19.9~$M_{Jup}$), and a radius of 0.35~$R_{\odot}$ (3.5$R_{Jup}$), also bigger than TWA~28 radius. \cite{Venuti2019} measured a luminosity of \textit{log~L} = -2.19~$L_{\odot}$ for TWA~27A. Similarly, as for TWA~28, TWA~27A also showed signs of a circumstellar disk due to near-infrared excess and emission lines in its X-shooter spectrum, measuring an accretion rate of $log(\dot{M}_{acc})$ = -11.23~$M_{\odot}\, yr^{-1}$.
In addition, using ALMA observations, \cite{Ricci2017} detected emission from dust continuum at 0.98~mm from the $J$ = 3 -- 2 rotational transition of the CO from a very compact disk around TWA~27A with a dust mass of approximate 0.1~$M_{\oplus}$. \cite{Ricci2017} argue that the small radius of the TWA~27A disk is due to the truncation due to the tidal interaction with its planetary-mass companion, TWA~27B.

TWA~27B is the first directly-imaged planetary-mass companion (M = 5$\pm$2~$M_{Jup}$, \citealt{Chauvin2005}). TWA~27A and b are separated by 0."77, corresponding to 40.6~au at a distance of 52.8~pc as estimated by \cite{Ducourant2008}. TWA~27B has a L6.0 spectral type and an effective temperature of 1300~K \citep{Luhman2023}. Its JWST/NIRSpec near-infrared spectrum shows emission lines suggesting the existence of a circumstellar disk \citep{Luhman2023}. Using ALMA observations, \cite{Ricci2017} estimated an upper limit of $\sim$1~$M_{Moon}$ for the mass of the dust surrounding TWA~27B.

{The coordinates, spectral types, magnitudes, distances and other fundamental parameters of our targets are summarized in Table \ref{table:parameters}.}

\begin{table*}
	\small
	\caption{Fundamental parameters of our targets.}  
	\label{table:parameters}
	\begin{center}
		\begin{tabular}{lllllllllll}
			\hline
			\hline 
				
			Name & RA  & DEC & $\mathrm{M_{J}}$ & SpT & log~g &  {Parallax (mas)} & Age (Myr) & Mass ($\mathrm{M_{Jup}}$) & \textit{log~L} ($L_{\odot}$) & References \\		
			\hline              
			TWA~28  & 11 02 09.83 & -34 30 35.56 & {13.0} & M9.0 & 4.10 & 16.87$\pm$0.13 & 10$\pm$2 & {20.9$\pm$6}   & -2.47 & 1, 2, {3} \\
            TWA~27A & 12 07 33.46 & -39 32 54.01 & {13.0} & M9.0 & 3.75 & 15.46$\pm$0.12 & 10$\pm$2 & {19.9$\pm$5}   & -2.19  & 1, 2, {3} \\
            TWA~27B & 12 07 33.50 & -39 32 54.40 & {19.5} & L6.0 & 3.50 & 15.46$\pm$0.12 & 10$\pm$2 & {5.0$\pm$2}     & -4.47 & 2, 3, 4\\
            \hline
				
		\end{tabular}
	\end{center}
	\begin{tablenotes}
		\small
		\item 1) \cite{Venuti2019}; 2) \cite{Gaia2020}; 3) \cite{Luhman2023}; 4) \cite{Chauvin2005}.\\
		
	\end{tablenotes}		
\end{table*}

\section{Observations} \label{sec:observations}

TWA~28, TWA~27A and TWA~27B were observed using {the} NIRSpec \citep{Jakobsen2022} Integral Field Unit (IFU, \citealt{Boeker2022}) onboard the \textit{James Webb Space Telescope} (JWST, \citealt{Gardner2023}), as part of the Guaranteed Time Observations (GTO) 1270 (P.I. Birkmann). The observations were obtained on February 7, 2023. The data were obtained using the medium-resolution gratings with {a resolution of R$\sim$2700 at $\sim$2.20~$\mu$m}, and a total wavelength coverage of 0.97--5.3~$\mu$m. The combination of gratings and filters used was: G140H/F100LP (0.97-1.89~$\mu$m), G235H/F170LP (1.66-3.17), and G395H/F290LP (2.87--5.27~$\mu$m). The observations were performed using the NRSIRS2RAPID readout pattern, with 32 groups for G140H/F100LP and G395H/F290LP, and 31 groups for G235H/F170LP, with 4 dithers. The total resulting observation time was $\sim$1900~s. The position angle of the telescope was designed to avoid bright stars in the field falling inside the  Multi-shutter assembly, which would contaminate the IFU spectra of our targets. Both TWA~27A and TWA~27B fitted inside the field of the IFU (3.1" X 3.2"). TWA28 was used to extract the spectrum of TWA~27B, since it has a similar spectral type to TWA~27A.
{The JWST data presented in this paper were obtained from the Mikulski Archive for Space
Telescopes (MAST) at the Space Telescope Science Institute. The specific observations
analyzed can be accessed via \dataset[DOI: 10.17909/gp39-v372]
{https://doi.org/10.17909/gp39-v372}.}

\section{Data Reduction} \label{sec:data_reduction}

\begin{figure*}[h]
    \centering
    \includegraphics[width=0.99\textwidth]{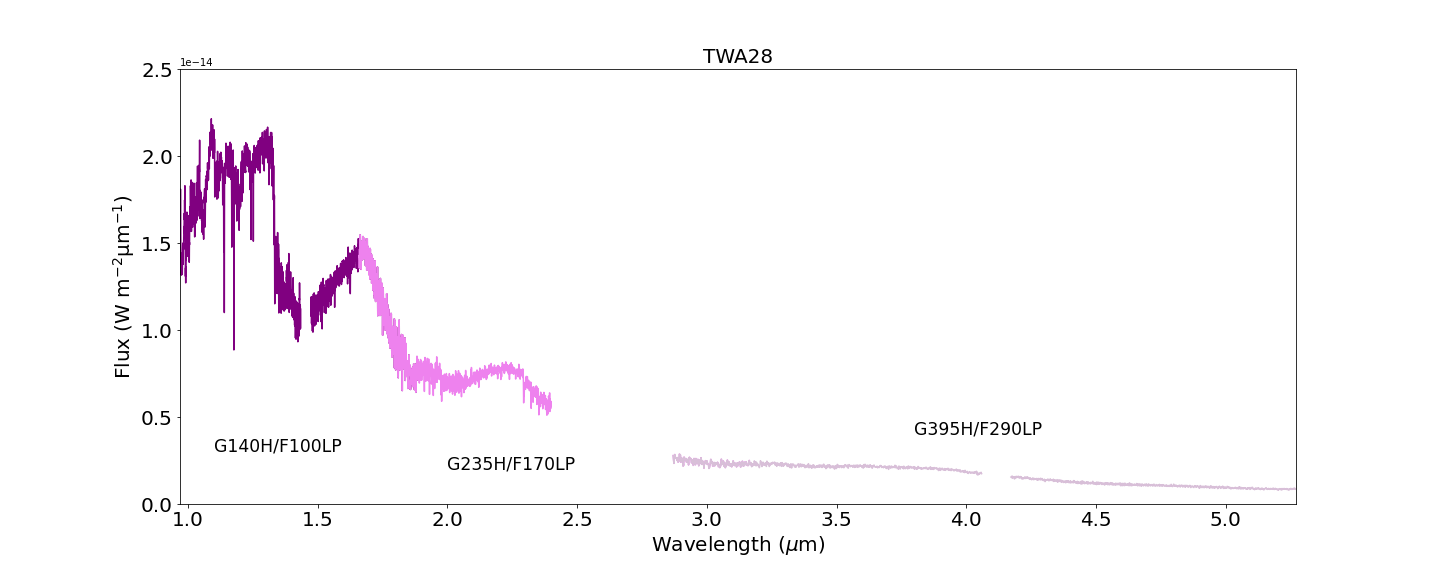}
    \includegraphics[width=0.99\textwidth]{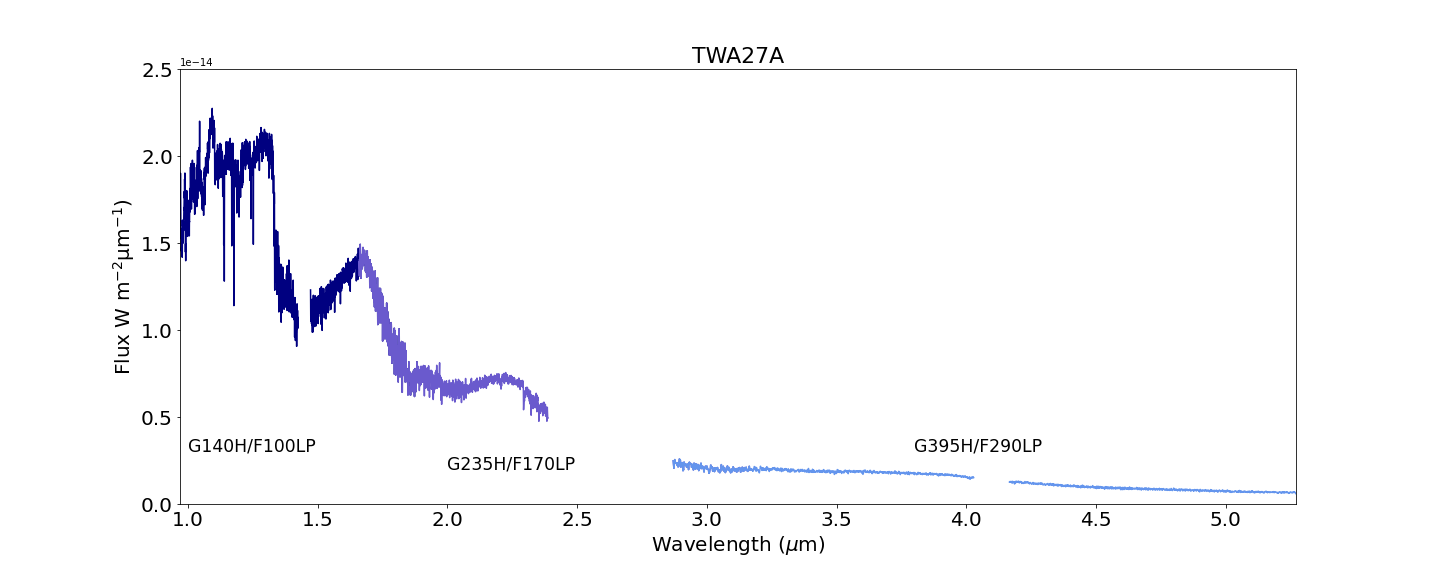}
    \includegraphics[width=0.99\textwidth]{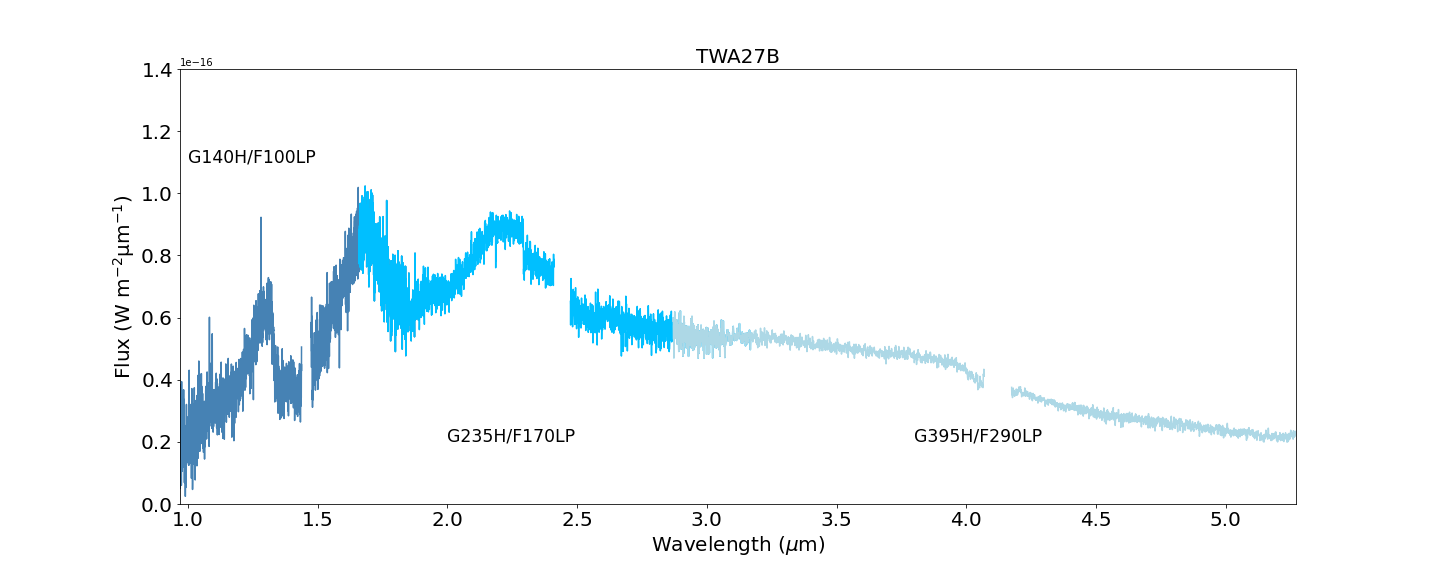}
   \caption{Full 0.97--5.3~$\mu$m NIRSpec/IFU {flux-calibrated spectra} of TWA 28, TWA~27A, and TWA~27B  respectively at their original resolution of R$\sim$2700. {We show in different colors the approximate wavelength ranges covered by each NIRSpec grating and filter combination.}}
    \label{fig:TWA2827Ab_full}
\end{figure*}

For the data reduction, we started with the ramps (‘uncal’ files in STScI pipeline notation) ‘raw’ data as downloaded from MAST and performed the ramps-to-slopes processing using the ESA Instrument Development Team pipeline, as it features a better correction for ‘snowballs’ and correction for residual correlated noise in IFU data. These count rate maps (‘rate’ files in STScI pipeline notation) were then used as inputs to the JWST \textit{calwebb\_spec2} pipeline {(version 1.9.4)} with the default processing steps, including assignment of WCS, flat fielding, and aperture correction. We then flagged outliers in the resulting ‘cal’ files using statistical methods, before building the IFU data cubes using the \textit{calwebb\_spec3} pipeline with outlier detection disabled/skipped. The latter was necessary as currently, the pipeline’s outlier detection is not working properly, resulting in many erroneously flagged data points and poor final cubes. The cubes were built with a spaxel size of 0.1” in the ‘ifualign’ coordinate system, in order to have the same PSF orientation between observations. Spectra for TWA~27A and TWA~28 were extracted from these cubes with an aperture radius of 4 spaxels, and we used the scaled and shifted TWA~28 cube to subtract from TWA~27A to get a contamination-free spectrum of TWA~27B.
We performed the same reduction steps on standard star P330E observed in program 1538 observation 62 with the NIRSpec IFU in the same gratings/filters and verified the flux calibration with the apertures used. {The minimum signal-to-noise of the reduced spectra for TWA~28 and TWA~27B is about $\sim$150, and $\sim$30 for TWA~27B. We need to note that the quality and signal-to-noise of the spectra might improve as the NIRSpec calibrations and the pipeline are improved in the upcoming cycles. The flux-calibrated reduced data are shown in Figure \ref{fig:TWA2827Ab_full}. In all the figures through the paper in which these spectra are shown, the spectrum of TWA~28 will be consistently shown in purple, the spectrum of TWA~27A will be shown in dark blue, and the spectrum of TWA~27B in light blue. }
The reduced spectra will be available in the electronic version of this article. 

\section{Data Analysis}\label{sec:data_analysis}

In this Section, we show the original resolution spectra of TWA~27A, TWA~27B, and TWA~28. We identify the most relevant atomic lines and molecular bands in each object. We performed a spectral indices analysis for the three objects to confirm their surface gravity. Finally, we will perform a modeling fitting to cloudy and cloudless {modeled spectra}.

\subsection{Identification of Spectral Lines and Molecular Bands}\label{sec:spectral_lines_bands}

{Together with the spectrum of VHS1256~b \citep{Miles2023}, the JWST/NIRSpec spectra presented in this paper for TWA~28, TWA~27A and TWA~27B have the highest signal-to-noise {ratio}, resolution, and {broadest} spectral coverage of all the available brown dwarfs and planetary-mass objects to date}. In this Section, we will visually identify the wealth of atomic and molecular lines identified in the spectra of these three young brown dwarfs/planetary-mass objects.
To be able to analyze in more detail the different spectral characteristics that appear in each object, we will zoom in on each wavelength range of interest. {For visualization purposes, we normalize all spectra to its median flux value in each range and apply an offset to show them together in the same plot}. We list the expected atomic lines and molecular bands as described in \cite{Cushing2005}. We also measure the equivalent width of the most prominent absorption lines, namely, the K\,I, Na\,I alkali lines. In addition, we will measure the prominent Paschen $\alpha$, $\beta$, $\gamma$ and $\delta$, together with the He\,I lines that are related to the potential existence of an accretion disk around TWA~27B.

\subsubsection{0.97--1.20~$\mu$m wavelength range}

\begin{figure*}[h]
    \centering
    \includegraphics[width=0.85\textwidth]{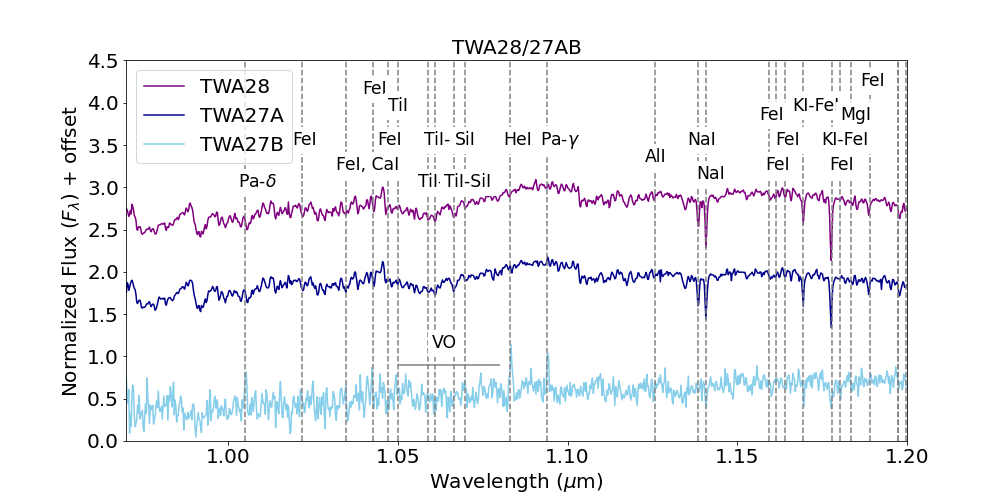}
    \includegraphics[width=0.85\textwidth]{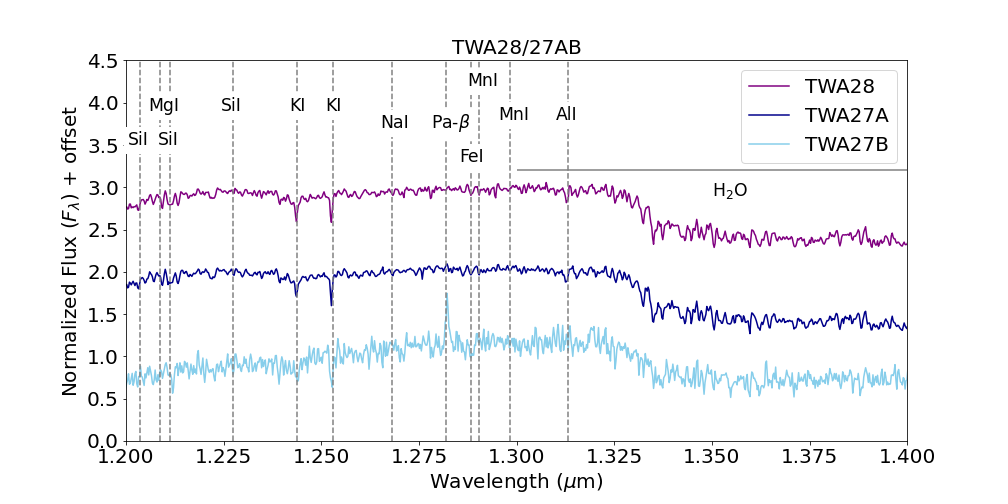}
    \includegraphics[width=0.85\textwidth]{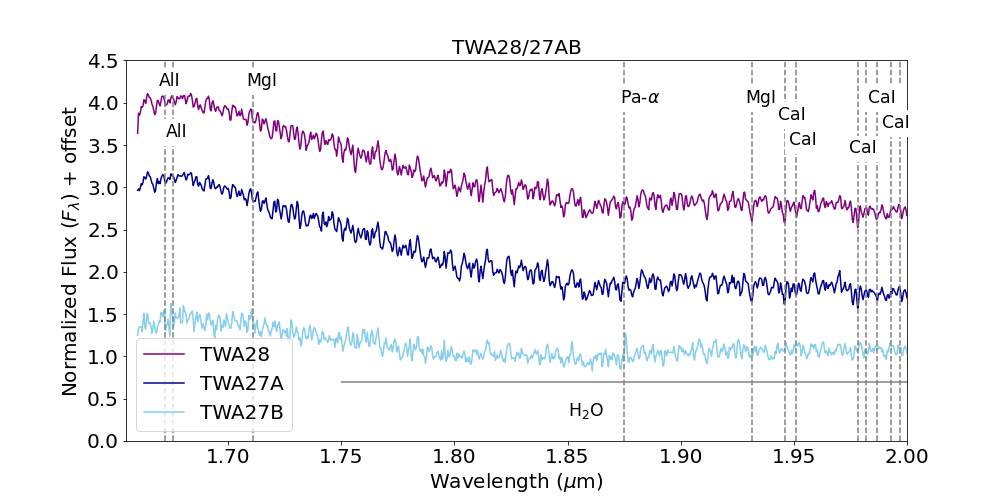}
   \caption{Full 0.97--2.0~$\mu$m NIRSpec/IFU spectrum of TWA 28 (purple), TWA~27A (dark blue), and TWA~27B (light blue) respectively at their original resolution of R$\sim$2700. We indicate the atomic absorption lines, emission lines and molecular bands we expect for these objects. }
    \label{fig:TWA2827Ab_1.0_2.0}
\end{figure*}

In the top panel of Fig. \ref{fig:TWA2827Ab_1.0_2.0}  we show the spectra of TWA~28, TWA~27A, and TWA~27B in the wavelength range between 0.97--1.20~$\mu$m. We marked the expected atomic lines, and molecular bands as listed in \cite{Cushing2005}, although not all of them are clearly detected. In TWA~28 and TWA~27A, the most prominent absorption lines are: the Na\,I doublet at 1.138 and 1.140~$\mu$m, the K\,I-Fe\,I doublet at 1.169 and 1.178~$\mu$m,  the Fe\,I line at 1.180~$\mu$m, the Fe\,I line at 1.0347~$\mu$m, and the Ti\,I--Si\,I line at 1.0664~$\mu$m.  Finally, we identify the VO molecular band between 1.05 and 1.08~$\mu$m.

The spectrum of TWA~27B has lower signal-to-noise, but we can still identify some prominent absorption lines: the Na\,I doublet at 1.138 and 1.140~$\mu$m, and the K\,I-Fe\,I doublet at 1.169 and 1.178~$\mu$m. For TWA~27B, the emission lines are quite prominent indicating the potential existence of an accretion disk, namely, these lines are the Paschen-$\delta$ line at 1.005~$\mu$m, the He\,I at 1.0830~$\mu$m, and the Paschen-$\gamma$ at 1.094~$\mu$m.

\subsubsection{1.20--1.40~$\mu$m wavelength range}

In the middle panel of Fig. \ref{fig:TWA2827Ab_1.0_2.0}  we show the spectra of TWA~28, TWA~27A, and TWA~27B in the wavelength range between 1.20--1.40~$\mu$m, with the expected atomic lines, and molecular bands as listed in \cite{Cushing2005}. In TWA~28 and TWA~27A, the most prominent absorption lines are: the K\,I doublet at 1.243~$\mu$m and 1.252~$\mu$m, the Al\,I line at 1.313~$\mu$m, the Na\,I line at 1.268~$\mu$m, and the Fe\,I line at 1.288~$\mu$m. Finally, from 1.3--1.51~$\mu$m, we identify the $\mathrm{H_{2}O}$ band.

As for the previous wavelength range, the spectrum of TWA~27B has lower signal-to-noise, but we can still identify the K\,I doublet at 1.243~$\mu$m and 1.252~$\mu$m. For emission lines, we identify a very prominent Paschen-$\beta$ line at 1.282~$\mu$m, and also the $\mathrm{H_{2}O}$ band from 1.3--1.51~$\mu$m.

\subsubsection{1.65--2.00~$\mu$m wavelength range}

In the bottom panel of Fig. \ref{fig:TWA2827Ab_1.0_2.0}  we show the spectra of TWA~28, TWA~27A, and TWA~27B in the wavelength range between 1.65--2.0~$\mu$m, with the expected atomic lines, and molecular bands as listed in \cite{Cushing2005}. In TWA~28 and TWA~27A, the most prominent absorption lines are: the Mg\,I line at 1.931~$\mu$m, and the Ca\,I lines at 1.986 and 1.993~$\mu$m. Finally, we detect the $\mathrm{H_{2}O}$ band from 1.75--2.05~$\mu$m.

For TWA~27B, the only clearly visible line is the Paschen-$\alpha$ line at 1.875~$\mu$m. The $\mathrm{H_{2}O}$ band from 1.75--2.05~$\mu$m is also present.

\subsubsection{2.00--2.39~$\mu$m wavelength range}

In the top panel of Fig. \ref{fig:TWA2827Ab_2.0_5.27}  we show the spectra of TWA~28, TWA~27A, and TWA~27B in the wavelength range between 2.0--2.39~$\mu$m. In TWA~28 and TWA~27A, the most prominent absorption lines are the Na\,I lines at 2.206 and 2.208~$\mu$m. No emission lines are detected for any of the targets in this wavelength range. Finally, we detect prominent CO-bands in TWA~28 and TWA~27A at 2.293, 2.322, 2.344, and 2.352~$\mu$m, but also the CO bands at 2.293 and 2.322~$\mu$m are visible in TWA~27B, although weaker than in the TWA~28 and TWA~27A spectra. 
This is to expect since TWA~27B is at the start of the L/T transition, where CO transforms in $\mathrm{CH_{4}}$ in the equilibrium reaction: $\mathrm{CO} + 3H_{2} \rightleftharpoons CH_{4} + H_{2}O$, thus CO should start depleting from TWA~27B's spectrum.

Nevertheless, if observe the comparison by \cite{Luhman2023} of TWA~27B with the spectrum of other young and intermediate age L6-L7 brown dwarfs (see their Fig. 5), namely CWISE~J050626.96+073842.4 (\citealt{Schneider2023}, $\sim$23~Myr), PSO~J318.5338--22.8603 (\citealt{Liu2013}, $\sim$23~Myr) and VHS~1256b ($\sim$140~Myr, \citealt{Dupuy2023}), we notice that the shape of the band is different, in particular between 2.10 and 2.30~$\mu$m, where the $\mathrm{CH_{4}}$ band stars to appear at 2.15~$\mu$m for L/T transition brown dwarfs \citep{Cushing2005}. In the spectrum of TWA~27B, no hints of the appearance of the $\mathrm{CH_{4}}$ band are seen.

\subsubsection{2.90--5.27~$\mu$m wavelength range}

\begin{figure*}[h]
    \centering
    \includegraphics[width=0.85\textwidth]{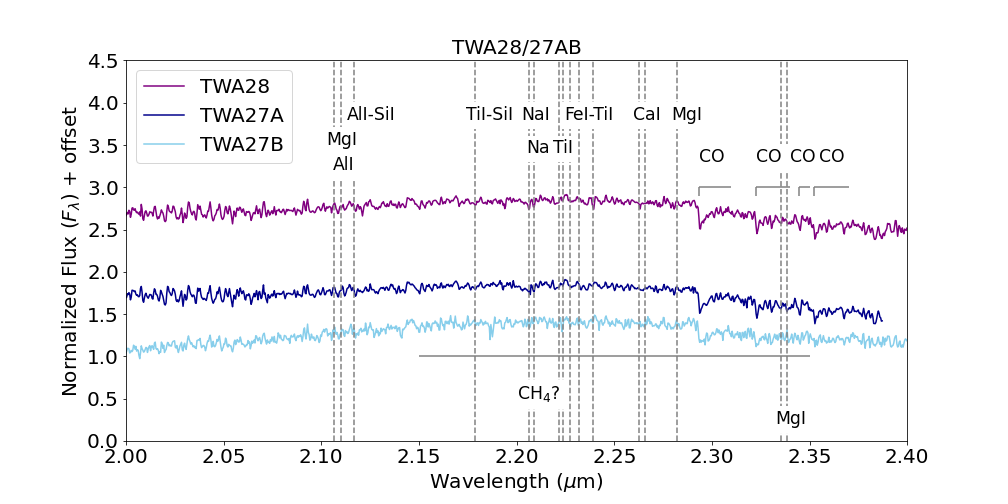}
    \includegraphics[width=0.85\textwidth]{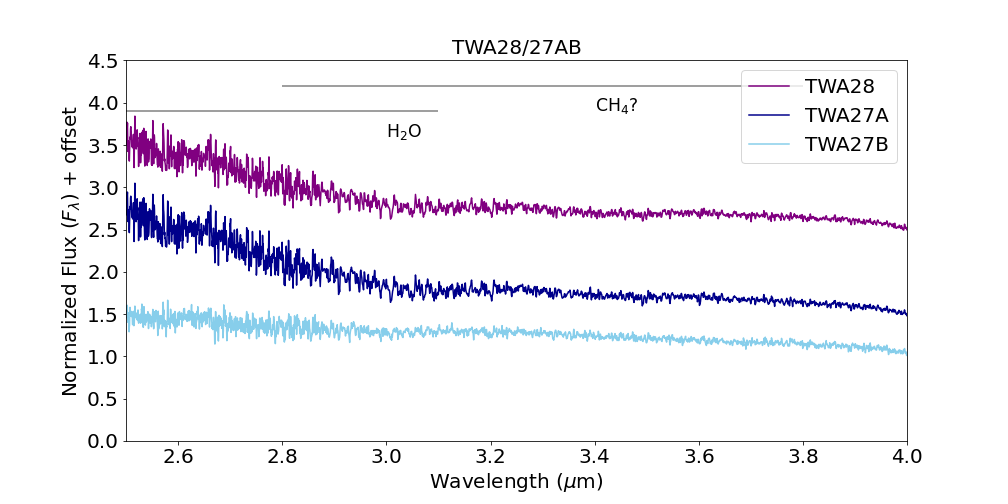}
    \includegraphics[width=0.85\textwidth]{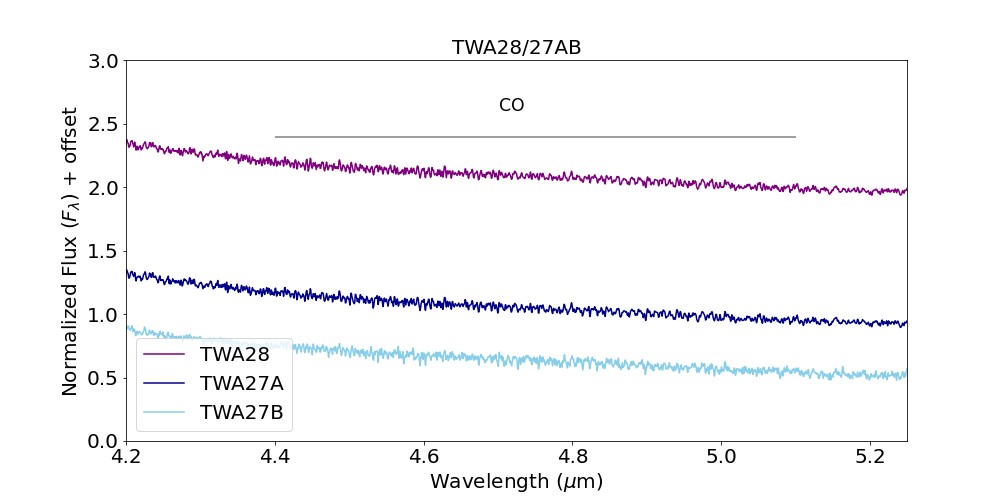}
   \caption{Full 2.0--5.27~$\mu$m NIRSpec/IFU spectrum of TWA 28 (purple), TWA~27A (dark blue), and TWA~27B (light blue) respectively at their original resolution of R$\sim$2700. We indicate the atomic absorption lines, emission lines and molecular bands we expect for these objects. }
    \label{fig:TWA2827Ab_2.0_5.27}
\end{figure*}

In Fig. \ref{fig:TWA2827Ab_2.0_5.27} we show the spectra of TWA 28 (purple), TWA~27A (dark blue), and TWA~27B (light blue) in the wavelength range between 2.9--5.27~$\mu$m. In this wavelength range, there are no {deep} atomic lines to highlight, but we do find a $\mathrm{H_{2}O}$ band from 2.5 to 3.1~$\mu$m, and a CO-band from 4.4--5.1~$\mu$m. Similarly as in the 2.00-2.39~$\mu$m range, since TWA~27B is a L6 brown dwarf, we would expect to detect $\mathrm{CH_{4}}$ between 2.8 and 3.8~$\mu$m as for VHS~1256b, and PSO-318, but this molecule is not present in TWA~27B spectrum as discussed in \cite{Luhman2023}. Similarly, the CO molecule for this object is weaker than for VHS~1256b, as also noted in \cite{Luhman2023}. \cite{Miles2023} also noticed that VHS~1256b has weaker $\mathrm{CH_{4}}$ between 2.8 and 3.8~$\mu$m than field brown dwarfs. The fact that both the CO and $\mathrm{CH_{4}}$ band are weaker in TWA~27B might be due to its lower surface gravity, as certainly some CO weakening has been observed at 2.4~$\mu$m for other young brown dwarfs \citep{Bonnefoy2014a, Manjavacas2014}. Alternatively, \cite{Miles2023} suggests that non-equilibrium chemistry would also explain the weakening of the $\mathrm{CH_{4}}$ band, although that should also strengthen the CO feature \citep{Mukherjee2022}.

\subsection{Emission Lines \& Circumstellar Disks}

Since our targets have been found to be accretors in previous works \citep{Venuti2019}, we searched for existing emission lines present in the NIRSpec spectrum further supporting this evidence. For TWA~27B, \cite{Luhman2023} showed the detection of emission of the $\alpha$, $\beta$, $\gamma$ and maybe $\delta$ lines, and the He\,I triplet at 1.083~$\mu$m, which are typical accretion signatures of young stars \citep{Natta2004}. \cite{Luhman2023} and \cite{Marleau2024} measured the accretion luminosity using the Pa$\beta$ line luminosity to derive an accretion rate of $\dot{M}$~=~$10^{-13}-10^{-12}M_{\odot}\, yr^{-1}$, which correlates with the accretion rates of young stars and brown dwarfs  derived by \cite{Muzerolle2005}  when extrapolating to the mass of TWA~27B. {The weak accretion rate implies that formation is likely over \citep{Marleau2024}}.

The presence of emission lines is not that evident for TWA~28 and TWA~27A in the NIRSpec spectra (see Fig. \ref{fig:TWA2827Ab_1.0_2.0}), even though both objects have been found to be accretors in previous works \citep{Venuti2019, Ricci2017}. The VLT/X-shooter spectra of TWA~28 and TWA~27A published by \cite{Venuti2019} also do not show emission lines in the near-infrared, but they do show emission lines in the ultraviolet and optical spectra, which allowed \cite{Venuti2019} to estimate their accretion rate. They estimated an accretion rate estimated for TWA~28 of $\mathrm{1.32\times10^{-5}}M_{\odot}\, yr^{-1}$, and  $\mathrm{4.83\times10^{-6}}M_{\odot} \,yr^{-1}$ for TWA~27A. The NIRSpec spectra of TWA~27A and TWA~28 show near-infrared color excess from $\sim$3~$\mu$m to 5.27~$\mu$m supporting the existence of a disk as reported previously \citep{Schneider2012, Venuti2019, Luhman2023}. {This color excess will be evident in Section \ref{sec:atmospheric_models} when we compare with atmospheric models}.

\subsection{Spectral Indices}\label{sec:spectral_indices}

We use the surface gravity spectral indices presented by \cite{Allers2013} to confirm the low surface gravity expected for TWA~28 and TWA~27A and b, given their membership to the TWA Hya association ($\sim$10~Myr, \citealt{Gizis2002}, \citealt{Mamajek2005}, \citealt{Weinberger}). {\cite{Allers2013} spectral indices were designed to measure the depth of different spectral characteristics in the near-infrared spectra of M5--L7 brown dwarfs that reflect their surface gravities. Using brown dwarfs with well-determined ages from 2~Myr to $\sim$1~Gyr, they provided a range of values for each of their indices and spectral types that would determine the surface gravity category of brown dwarfs (see their Table~9). Those three categories are field (FLD-G, gravity score 0, {log $g$ = 5.0-5.5}), intermediate-gravity (INT-G, gravity score 1, {log $g$ = 4.0-4.5}), and very-low gravity (VL-G, gravity score 2, {log $g$ $<$ 4.0})}.

\begin{figure}[h]
    \centering
    \includegraphics[width=0.5\textwidth]{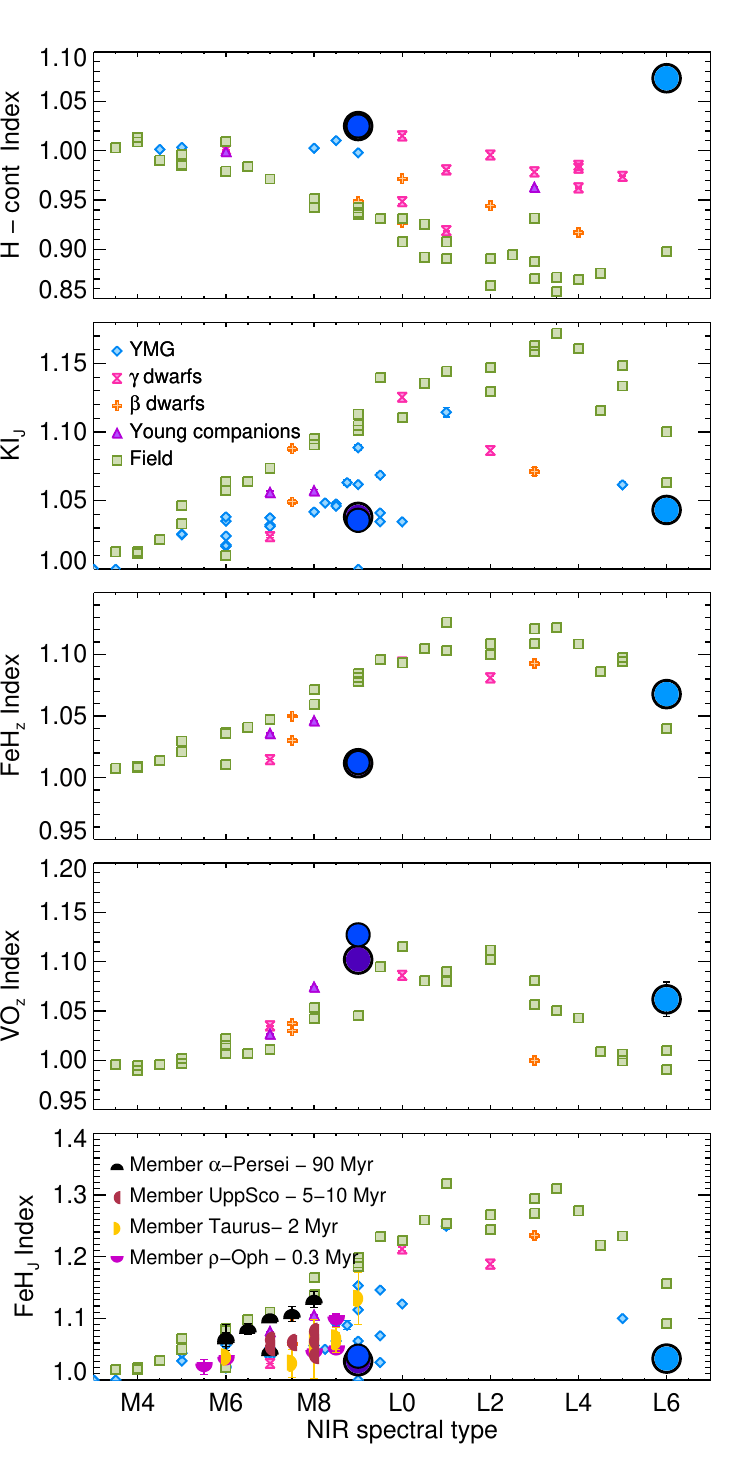}
   \caption{Gravity scores from \cite{Allers2013} obtained for our targets {(purple circle = TWA~28, smaller dark blue circle = TWA~27A, light blue circle = TWA~27B)}, and for other M3-L7 objects with different ages as a comparison. We show for comparison: field objects \citep{McLean, Cushing2005}, young companion \citep{Allers2007, Bonnefoy2014a}, young $\beta$ and $\gamma$ dwarfs. For members of the $\alpha$ Persei, Upper Scorpious, Taurus, and $\rho$ Ophiuchi, we include the value of the FeH index \citep{Martin2017}.}
    \label{gravity_indices}
\end{figure}

\begin{figure}[h]
    \centering
    \includegraphics[width=0.5\textwidth]{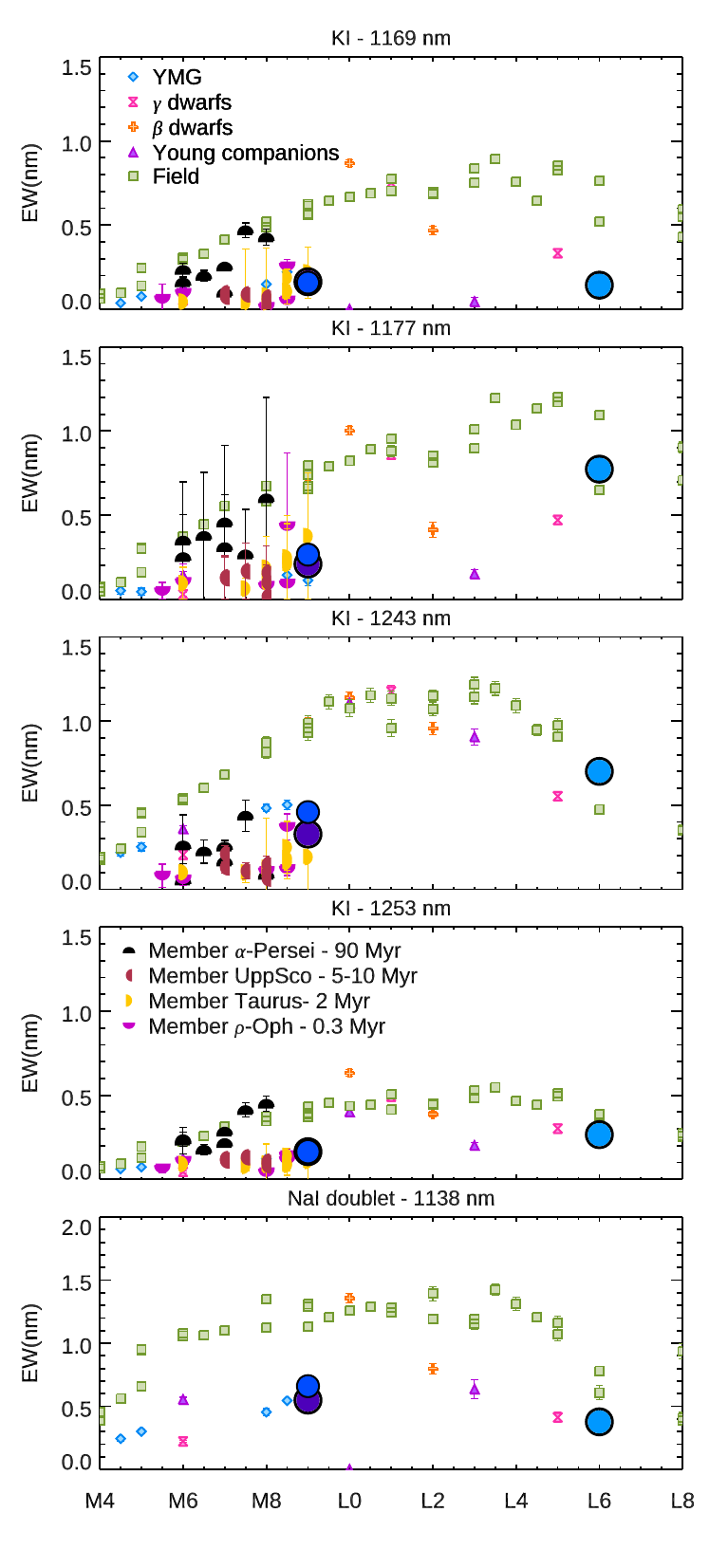}
   \caption{Equivalent widths of the K\,I and Na\,I alkali lines for our targets {(purple circle = TWA~28, smaller dark blue circle = TWA~27A, light blue circle = TWA~27B)} and the same comparison objects mentioned in Fig. \ref{gravity_indices}.}
    \label{pEW}
\end{figure}

We calculated the $\mathrm{FeH_{J}}$, $\mathrm{KI_{J}}$, $\mathrm{FeH_{z}}$, $\mathrm{VO_{z}}$, and $H$-cont spectral indices {presented in \cite{Allers2013}. The specific wavelength ranges that each of these indices span, and how the final index values are obtained are shown in Table~4 and Equation~1 in \cite{Allers2013}}. The $\mathrm{FeH_{J}}$ is correlated with surface gravity \citep{McGovern}, the $\mathrm{KI_{J}}$ index measures the pseudo equivalent widths of the K\,I doublet in the $J$-band at 1.250~$\mu$m. The VO band at 1.058~$\mu$m disappears with the increment of surface gravity \citep{Lodders2002}. The $H$-cont index measures the shape of the $H$-band, which has been found to become more triangular with low surface gravity \citep{Bowler2012}. \cite{Lodieu2017} identified the $H$-cont index as the most sensitive to surface gravity. In Fig. \ref{gravity_indices}, we compared the values of these indices for TWA~28, TWA~27A and TWA~27B with those {measured for other M4-L6 low-mass stars and brown dwarfs with some indication of age: } young moving groups (YMG), $\gamma$ {(low surface gravity, log $g$ $<$ 4.0)} and $\beta$ dwarfs {(intermediate surface gravity, log $g$ = 4.0-4.5)}\citep{Allers2007, Bonnefoy2014a}, young companions \citep{Allers2007}, and field brown dwarfs \citep{McLean, Cushing2005} with spectral types between M4 and L8. We also compared with members of different open clusters and associations, from \cite{Martin2017}, namely: $\alpha$ Persei (90~Myr), Upper Scorpious (5--10~Myr), Taurus (2~Myr), and $\rho$ Ophiucus (0.3~Myr).  In addition, alkali lines have been identified as surface gravity indicators for brown dwarfs. For young brown dwarfs, the alkali lines have been observed to be weaker than for field brown dwarfs \citep{Steele_Jameson1995, Martin1996,Gorlova2003,Cushing2005,Allers2007, Allers2013,Bonnefoy2014a, Manjavacas2014}. We measured the pseudo equivalent widths of the K\,I lines at 1.169, 1.177, 1.243 and 1.253~$\mu$m, and the Na\,I doublet at 1.138~$\mu$m (see Fig. \ref{pEW}), {using the same wavelength ranges than in \cite{Allers2013}, their Table~7}. Similarly, as we did with the spectral indices, we compared the values of the pseudo-equivalent widths with those of other brown dwarfs of different surface gravities from the literature.  In Table \ref{table:ew_all_lines_nir} we show the value of the pseudo equivalent widths of the alkali lines for TWA~28, TWA~27A and TWA~27B, and in Table \ref{table:gscores} we show the gravity scores for the three objects. As expected for the young age of the objects ($\sim$10~Myr), we obtained that the majority of the gravity indices classified these objects as very-low surface gravity objects (VL-G). 

	\begin{table*}
		\small
		\caption{Equivalent widths in nm for alkali lines measured in the near-infrared.}  
		\label{table:ew_all_lines_nir}
		\centering
		\begin{center}
			\begin{tabular}{llllllll}
				\hline
				\hline 
				
				Name & NIR SpT  & K\,I~(1.169~$\mu$m)  & K\,I~(1.177~$\mu$m) & K\,I~(1.243~$\mu$m) & K\,I~(1.253~$\mu$m) & Na\,I (1.138~$\mu$m) & GS $\mathrm{^{a, b}}$\\		
				\hline              
				TWA~28  & M9.0   & 0.15$\pm$0.02 & 0.27$\pm$0.03   & 0.46$\pm$0.02   & 0.16$\pm$0.01   & 0.66$\pm$0.03 &  22-22 /  VL-G  \\
                TWA~27A & M9.0   & 0.16$\pm$0.01   & 0.21$\pm$0.03   &0.33$\pm$0.02    & 0.16$\pm$0.01  & 0.55$\pm$0.02  &    22-12    / VL-G     \\
                TWA~27B & L6.0    & 0.14$\pm$0.07   & 0.77$\pm$0.06   &0.70$\pm$0.08    & 0.27$\pm$0.04 & 0.38$\pm$0.11    &    21-22    / VL-G    \\
                \hline
				
			\end{tabular}
		\end{center}
				\begin{tablenotes}
		\small
		\item a: GS: Gravity scores calculated as in \cite{Allers2013}. b: Gravity scores are ordered according to the alkali line that they correspond to. The dash symbol indicates that none gravity score can be determined with that particular line.\\
		
	\end{tablenotes}		
	\end{table*}

 	\begin{table*}
		\small
		\caption{Indices values and gravity scores for our sample derived from spectral indices defined in the literature \citep{Allers&Liu}.}  
		\label{table:gscores}
		\centering
		\begin{center}
			\begin{tabular}{llllllll}
				\hline
				\hline 
				
				Name     &  SpT &$\mathrm{FeH_{J}}$ & $\mathrm{KI_{J}}$ & $H$-cont          &  $\mathrm{FeH_{z}}$  &  $\mathrm{VO_{z}}$ & G$\mathrm{S^{a}}$  \\		
				\hline              
				TWA~28    & M9.0 & 1.039$\pm$0.001	&   1.036$\pm$0.001 & 1.024$\pm$0.001 &	1.012$\pm$0.002 & 1.102$\pm$0.002       & 2221 --/ VL-G \\
                TWA~27A    & M9.0 & 1.030$\pm$0.001	&   1.038$\pm$0.001 & 1.024$\pm$0.002	 &	1.011$\pm$0.001 & 1.100$\pm$0.002		& 2222 -- / VL-G   \\
                TWA~27B    & L6.0 & 1.034$\pm$0.001	&   1.043$\pm$0.001 & 1.073$\pm$0.006	 &	1.068$\pm$0.001 & 1.062$\pm$0.002		& 2--2-- -- / VL-G  \\

                                \hline
			\end{tabular}
		\end{center}		
		\begin{tablenotes}
		\small
		\item a: GS: Gravity scores calculated as in \cite{Allers2013}\\		
	\end{tablenotes}		
	\end{table*}

	\begin{table*}
		\small
		\caption{Equivalent widths in nm for the Paschen series emission lines and He\,I measured in the near-infrared.}  
		\label{table:ew_emission_lines_nir}
		\centering
		\begin{center}
			\begin{tabular}{lllllll}
				\hline
				\hline 
				
				Name & NIR SpT  & Paschen $\alpha$  & Paschen $\beta$ & Paschen $\gamma$ & Paschen $\delta$ & He\,I \\		
				\hline              
                TWA~27B & L6.0    & -0.350$\pm$0.028   & -0.425$\pm$0.026   & -0.504$\pm$0.047    & -0.745$\pm$0.084 & -0.296$\pm$0.099      \\
                \hline
				
			\end{tabular}
		\end{center}		

 	\end{table*}

\section{Comparison with Atmospheric Models}\label{sec:atmospheric_models}

We compare the spectra of TWA~28, TWA~27A and TWA~27B at its original resolution (R$\sim$2700) to different set of atmospheric models to derive their effective temperatures, surface gravities, and other atmospheric properties like the radius and mass of the objects. For this purpose we use two types of models, the {\tt BT-Settl} models \citep{Allard2012a}, which include cloud implementation \citep{Allard2003,Allard2012a}, and with the {\tt ATMO} cloudless models \citep{Tremblin2015, Tremblin2017}. We used the software {\tt species}\footnote{\url{https://species.readthedocs.io/en/latest/index.html}} \citep{Stolker2020} to fit the {\tt BT-Settl} models to our NIRSpec spectra. We fit only the 0.97--2.40~$\mu$m, since the three objects have circumstellar disks, which affect mostly the SED from 2.40~$\mu$m to 5.30~$\mu$m. 



For the {\tt BT-Settl} models \citep{Allard2012a} the cloud model is implemented in the {\tt PHOENIX} atmosphere code \citep{Allard2001}. All relevant molecular absorbers are treated with line-by-line opacities in direct opacity sampling as in \cite{Allard2003}. The {\tt BT-Settl} models include disequilibrium chemistry for CO, $\mathrm{CH_{4}}$, $\mathrm{CO_{2}}$, $\mathrm{N_{2}}$, and $\mathrm{NH_{3}}$. We use the most updated version of the {\tt BT-Settl} models which are based on the {\tt CIFIST} photospheric solar abundances \citep{Caffau2011} to compare to TWA~28 and TWA~27A, and the previous version of the {\tt BT-Settl} models to compare to TWA~27B, since it showed a better match with its spectrum.

The cloudless {\tt ATMO} models \citep{Tremblin2015, Tremblin2017} propose that fingering convection is responsible for this turbulent mixing and leads to the out-of-equilibrium abundances in the atmospheres of brown dwarfs. Fingering convection can be triggered in the atmospheres of brown dwarfs because of the gradient of mean molecular weight induced by the chemical transitions CO/$\mathrm{CH_{4}}$ and $\mathrm{N_{2}/NH_{3}}$ at chemical equilibrium \citep{Tremblin2017}. Fingering convention reduces the temperature gradient reproducing the reddening of the spectra of young brown dwarfs \citep{Tremblin2017}.

In this Section, we compare the atmospheric parameters derived using both models for TWA~28 and TWA~27A and TWA~27B at a resolution of R$\sim$2700. We use the software {\tt species} {to find the best matching synthetic spectra generated using the {\tt BT-Settl} models}. {The toolkit  {\tt species} {is designed} to analyze spectral and photometric data {of} planetary and substellar atmospheres. It} uses a Monte Carlo algorithm to sample the parameter space called {\tt PyMultiNest}\footnote{\url{http://johannesbuchner.github.io/PyMultiNest/index.html}} with 2000 live points. {The toolkit {\tt species} also derives the radius (R, in $\mathrm{R_{Jup}}$), and mass (M, in $\mathrm{M_{Jup}}$) using the trigonometric parallax provided ($\omega$, mas), and the $\mathrm{T_{eff}}$ and log~$g$ of the best matching synthetic spectrum. We derived the radius of the targets by using the dilution factor $\mathrm{C = (R/d)^{2}}$, which takes into account the flux dilution between the synthetic spectra generated at the outer boundary of the atmosphere and the flux-calibrated spectrum. R is the brown dwarf radius and d is the distance, calculated using the parallax. 
The mass is derived using the estimated surface gravity with the model (log~g) using the expression: $\mathrm{g = G M/R^{2}}$, where G is the gravitational constant, from which M can be obtained}. 

{We double-check all the best matches obtained with {\tt species} by matching the atmospheric grids to the spectra using a $\chi^{2}$ fitting method. Both methods provided consistent best matches for the three objects.}

We will describe the divergences between both models and the spectra in the overall SED (see Figure \ref{best_fit_models-TWA28_TWA27A_TWA27B}), and also in the individual wavelength ranges which will allow us to describe in more detail smaller discrepancies. This comparison can be found  in Figures \ref{TWA28_BT-Settl+nodisk_ATOM_0.95-2.0} through \ref{TWA27b_BT-Settl+nodisk_ATOM_2.0-5.2}. Note that the uncertainties derived from PyMultiNest are just statistical and derived from the propagation of the small error bars of our data through the Bayesian inversion, thus, the uncertainties of the derived parameters are likely underestimated. {More conservative uncertainties for the effective temperature are $\pm$100~K, and $\pm$0.5 in \textit{log~g}, since that is the resolution of the grid. We derived the uncertainties in Table~\ref{table:summary_results} using those uncertainties. Due to the uncertainty in log~g, because of the width of the grid, the uncertainties in the mass might be likely overestimated.}

\subsection{TWA~28 Model Comparison}

Using {\tt species}, we find the best match to the TWA~28 JWST/NIRSpec spectrum using the {\tt BT-Settl CIFIST} atmospheric models. The spectrum of TWA~28 is best reproduced using a model with effective temperature ($\mathrm{T_{eff}}$) of 2577~K, a surface gravity (\textit{log g}) of 4.0, a radii (\textit{R}) of 2.54~$\mathrm{R_{Jup}}$, and a mass for TWA~28 of 26.0~$\mathrm{M_{Jup}}$.  
In the Appendix, we also show the best-fit parameters derived by {\tt species} in Figure \ref{multinest-TWA28}.

\begin{figure*}[h]
\centering
    \includegraphics[width=0.7\textwidth]{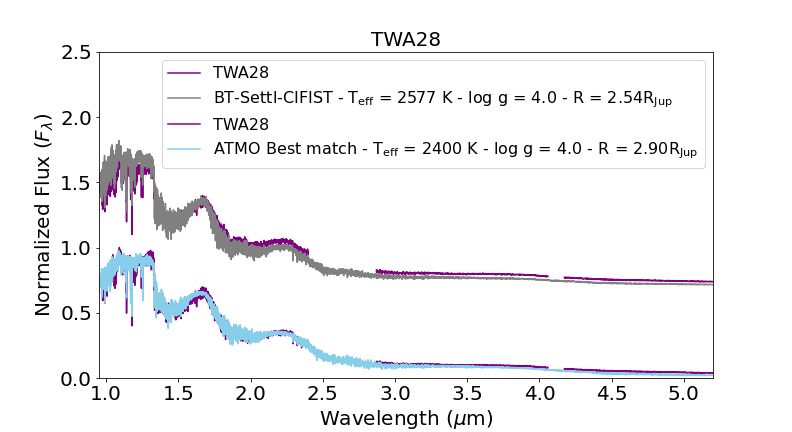}
    \includegraphics[width=0.7\textwidth]{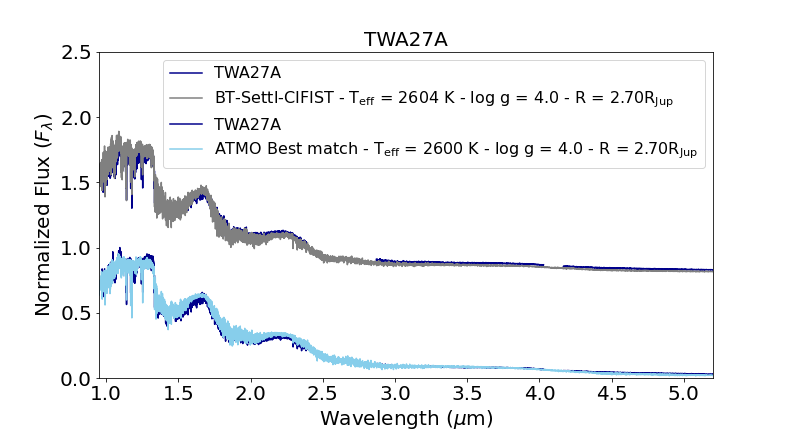}
    \includegraphics[width=0.7\textwidth]{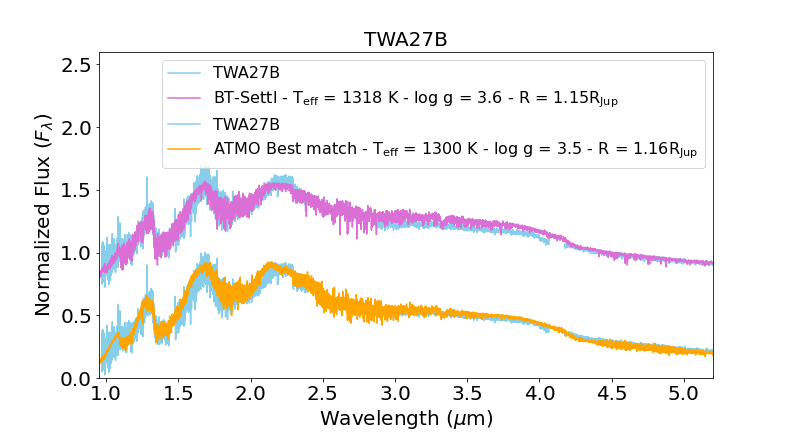}
   \caption{Best matches found for the 0.97-5.3~$\mu$m spectra of TWA~28 {(purple)}, TWA~27A {(dark blue)} and TWA~27B {(light blue)} using the {\tt BT-Settl} model,  and the {\tt ATMO} models. {}}
    \label{best_fit_models-TWA28_TWA27A_TWA27B}
\end{figure*}

In the overall SED, the {\tt BT-Settl CIFIST} models over-predict the flux in the 0.97--1.30~$\mu$m wavelength range, but under-predict the flux in the 2.00--2.50~$\mu$m range. After 2.5~$\mu$m, the contribution to the Spectral Energy Distribution (SED) of TWA~28 starts to be affected by the contribution of the circumstellar disk, thus, the underpredicted flux observed on the redder part of the spectrum is to be expected (see Fig. \ref{best_fit_models-TWA28_TWA27A_TWA27B}).

The {\tt ATMO} models provide a {colder} effective temperature of $\mathrm{T_{eff}}$ = 2400~K, but similar \textit{log~g}=4.0 and a sligher bigger radius of R = 0.3~$\mathrm{R_{\odot}}$ or 2.90~$\mathrm{R_{Jup}}$, and a mass of 32.4~$\mathrm{M_{Jup}}$. The {\tt ATMO} models provide a satisfactory fit to the 0.97--2.40~$\mu$m wavelength range (see Fig. \ref{best_fit_models-TWA28_TWA27A_TWA27B}). Since no circumstellar disk is included in the fit, from 2.5~$\mu$m on the fit starts to diverge as expected.
{A summary of the fundamental physical parameters derived for TWA~28 are summarized in Table \ref{table:summary_results}}.

\begin{figure*}
    \centering
    \includegraphics[width=0.85\textwidth]{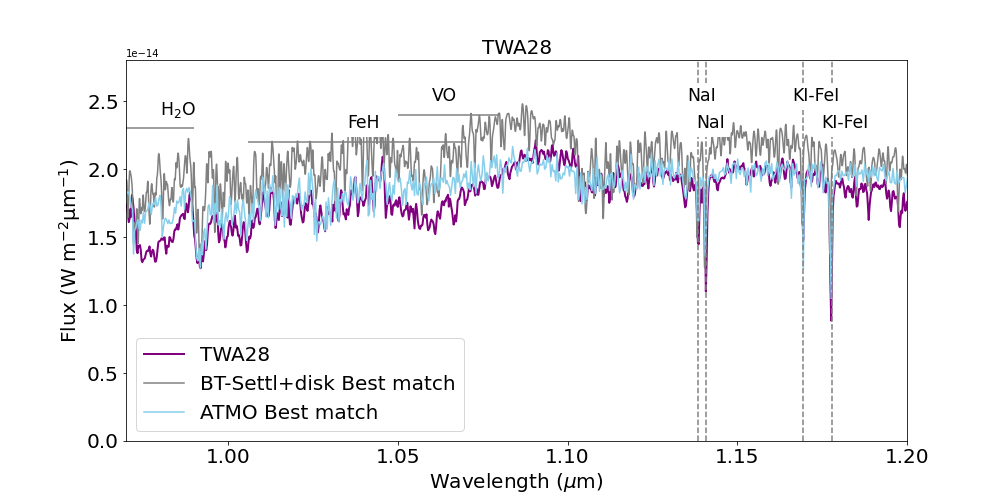}
    \includegraphics[width=0.85\textwidth]{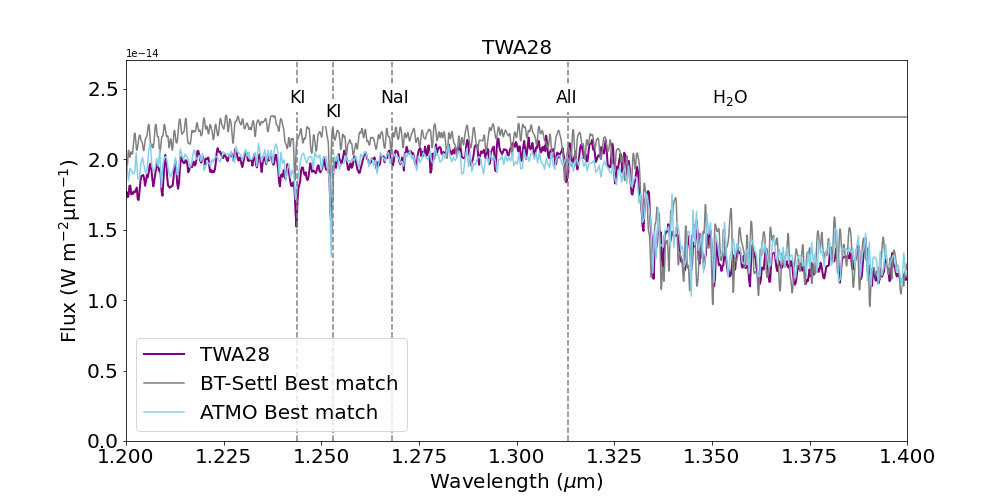}
    \includegraphics[width=0.85\textwidth]{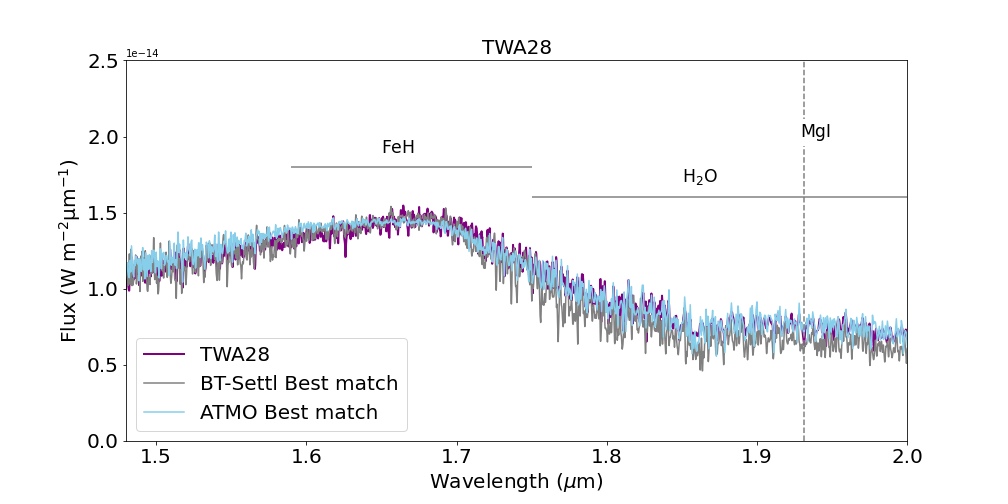}
   \caption{Best fit of the {\tt ATMO} and {\tt BT-Settl CIFIST} models to the TWA~28 0.97-2.0~$\mu$m spectrum. We highlight the main atomic lines and molecular absorptions present in this wavelength range.}
    \label{TWA28_BT-Settl+nodisk_ATOM_0.95-2.0}
\end{figure*}

\begin{figure*}
    \centering
    \includegraphics[width=0.85\textwidth]{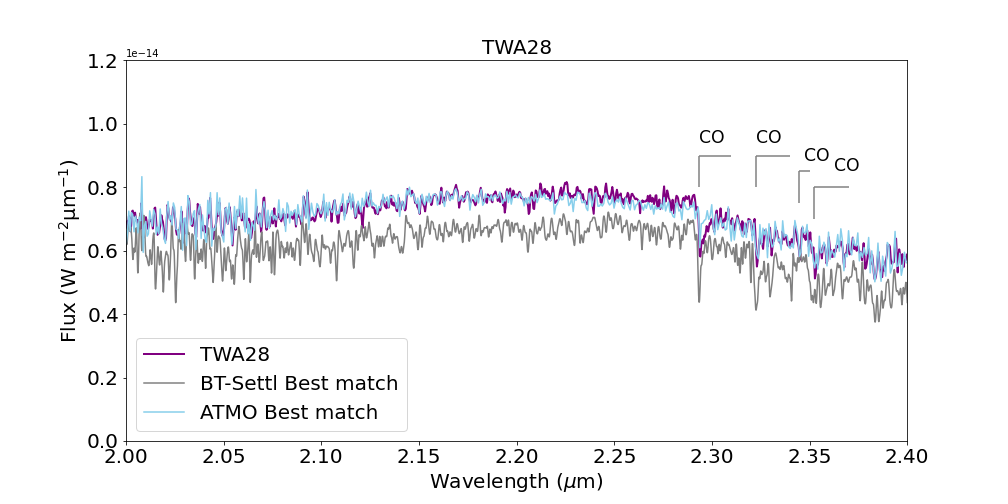}
    \includegraphics[width=0.85\textwidth]{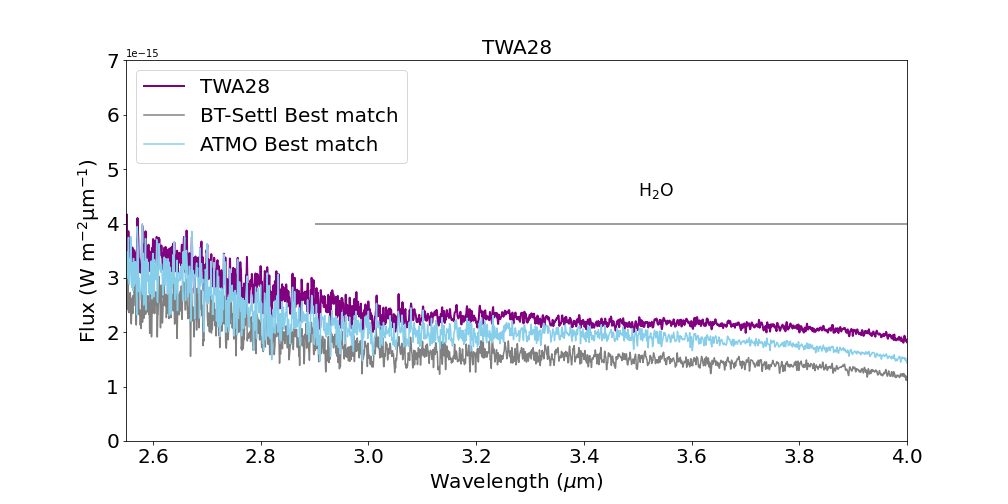}
    \includegraphics[width=0.85\textwidth]{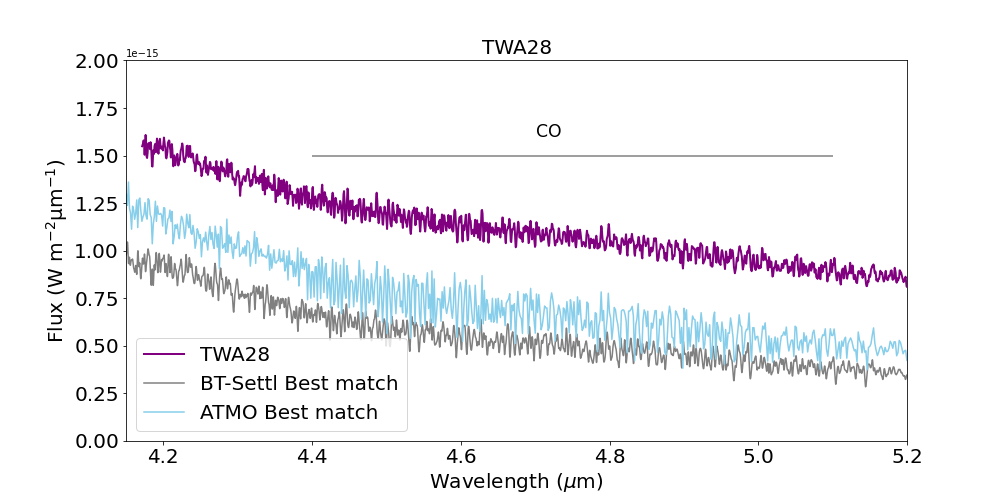}
   \caption{Best fit of the {\tt ATMO} and {\tt BT-Settl CIFIST} models to the TWA~28 2.0-5.20~$\mu$m spectrum. We highlight the main molecular absorptions present in this wavelength range.}
    \label{TWA28_BT-Settl+nodisk_ATOM_2.0-5.2}
\end{figure*}

Comparing the individual wavelength ranges as shown in Fig. \ref{TWA28_BT-Settl+nodisk_ATOM_0.95-2.0} and \ref{TWA28_BT-Settl+nodisk_ATOM_2.0-5.2}, we notice few characteristics that the models do not fully reproduce: in the 0.97-1.20~$\mu$m range, neither of the models satisfactorily reproduce the water band at 0.98~$\mu$m or the FeH or VO bands at 1.04 and 1.06~$\mu$m, respectively. The Na\,I and K\,I alkali lines are not as deep in the models as in the NIRSpec spectrum. In the 1.20 to 1.40~$\mu$m range, the K\,I alkali line at 1.243~$\mu$m is too weak, while the K\,I line at 1.253~$\mu$m is too strong in comparison with the NIRSpec spectrum. The wavelength range between 1.48 and 2.0~$\mu$m is well reproduced by the {\tt BT-Settl CIFIST} model, while the {\tt ATMO} model over-predicts the flux in the 1.48-1.65~$\mu$m region,  not reproducing accurately the typical triangular shape of this band for young brown dwarfs. In the 2.00-2.40~$\mu$m region, the {\tt BT-Settl} CIFIST model does not reproduce exactly the CO bands between 2.30-2.40~$\mu$m.  In the 4.0-5.2~$\mu$m region, we notice that the {\tt ATMO} models predict {deeper} CO bands in comparison to the CO bands observed in the NIRSpec spectrum, and the prediction of the {\tt BT-Settl CIFIST} models.

\subsection{TWA~27A Model Comparison}

Similarly as for TWA~28, we use {\tt species} to find the best match to the JWST/NIRSpec TWA~27A spectra using the {\tt BT-Settl CIFIST} atmospheric models. We obtained an effective temperature of $\mathrm{T_{eff}}$ = 2605~K, a \textit{log~g} = 4.0, a radius of R = 2.70~$\mathrm{R_{Jup}}$ and a mass of 29.3~$\mathrm{M_{Jup}}$. 
The {\tt BT-Settl} CIFIST models over-predict the flux in the 0.97--1.30~$\mu$m wavelength range, and under-predicts the flux in the 2.00--2.40~$\mu$m. (see Fig. \ref{best_fit_models-TWA28_TWA27A_TWA27B}). Similarly to TWA~28, since no circumstellar disk is included in the fit, from 2.50~$\mu$m on the fit starts to diverge (see Fig. \ref{best_fit_models-TWA28_TWA27A_TWA27B}. In the Appendix, we also show the best-fit parameters derived by {\tt species} in Figure \ref{multinest-TWA27A}.

The {\tt ATMO} models provide a similar effective temperature of $\mathrm{T_{eff}}$ = 2600~K,  similar \textit{log~g} = 4.0 and a sligher bigger radius of R = 0.285~$\mathrm{R_{\odot}}$ or 2.70~$\mathrm{R_{Jup}}$, and a mass of 28.1~$\mathrm{M_{Jup}}$. The {\tt ATMO} models provide a satisfactory fit in the 0.97--2.40~$\mu$m wavelength range, but does no reproduce the triangular shape of the 1.50-1.70~$\mu$m wavelength range, characteristic of young brown dwarfs. Similarly to TWA~28, no circumstellar disk is included in the fit, and from 2.50~$\mu$m on the fit starts to diverge (see Fig. \ref{best_fit_models-TWA28_TWA27A_TWA27B}). 
{A summary of the fundamental physical parameters derived for TWA~27A are summarized in Table \ref{table:summary_results}}.

\begin{figure*}
    \centering
    \includegraphics[width=0.85\textwidth]{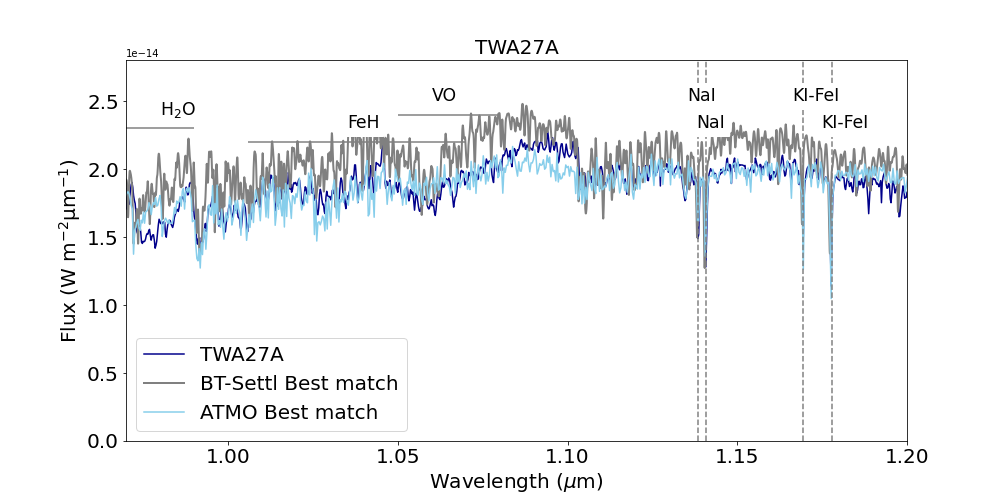}
    \includegraphics[width=0.85\textwidth]{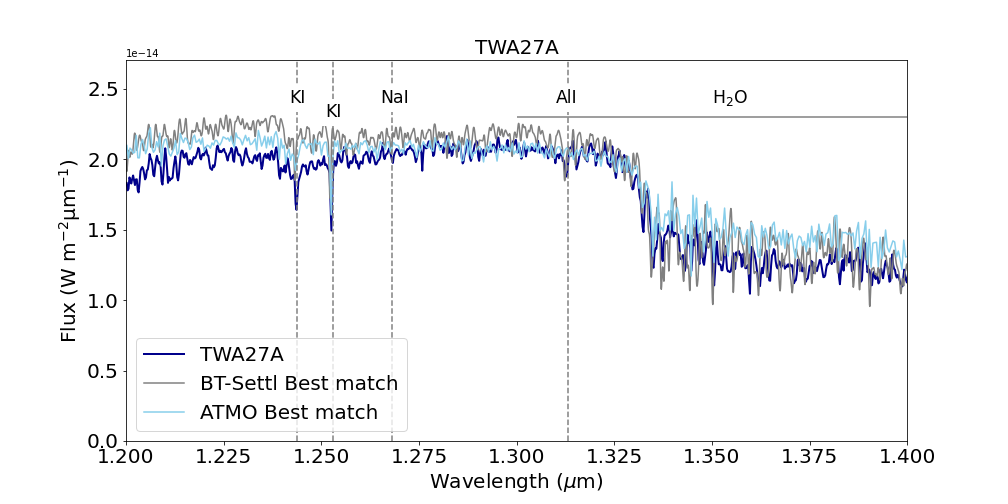}
    \includegraphics[width=0.85\textwidth]{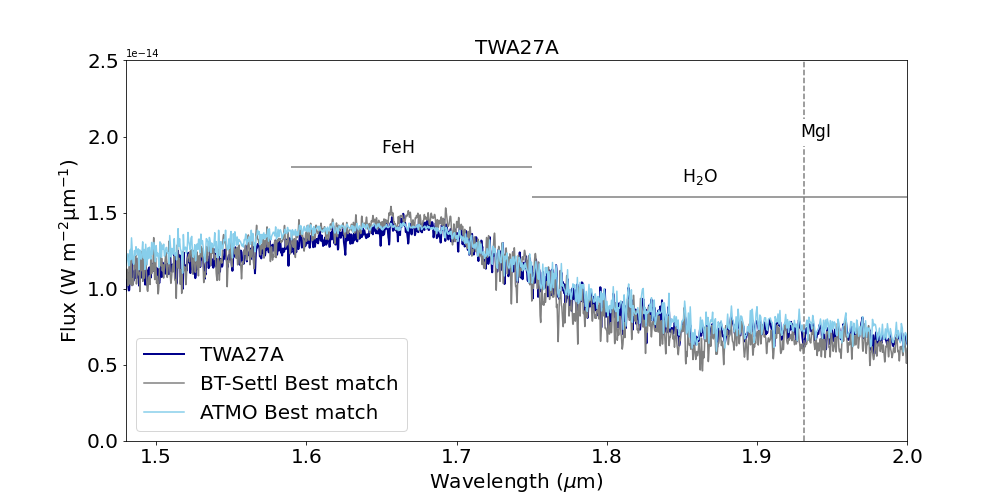}
   \caption{Best fit of the {\tt ATMO} and {\tt BT-Settl CIFIST} models to the TWA~27A 0.97-2.0~$\mu$m spectrum. We highlight the main atomic lines and molecular absorptions present in this wavelength range.}
    \label{TWA27A_BT-Settl+nodisk_ATOM_0.95-2.0}
\end{figure*}

\begin{figure*}
    \centering
    \includegraphics[width=0.85\textwidth]{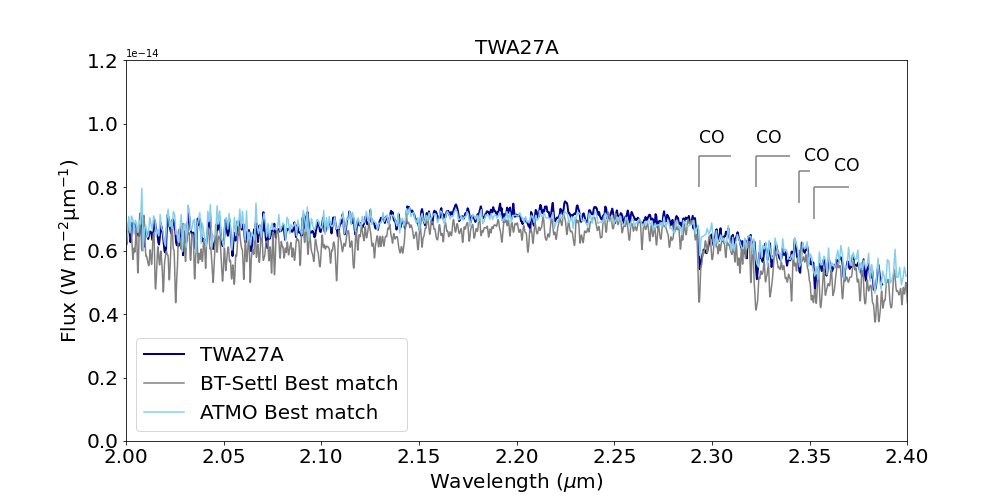}
    \includegraphics[width=0.85\textwidth]{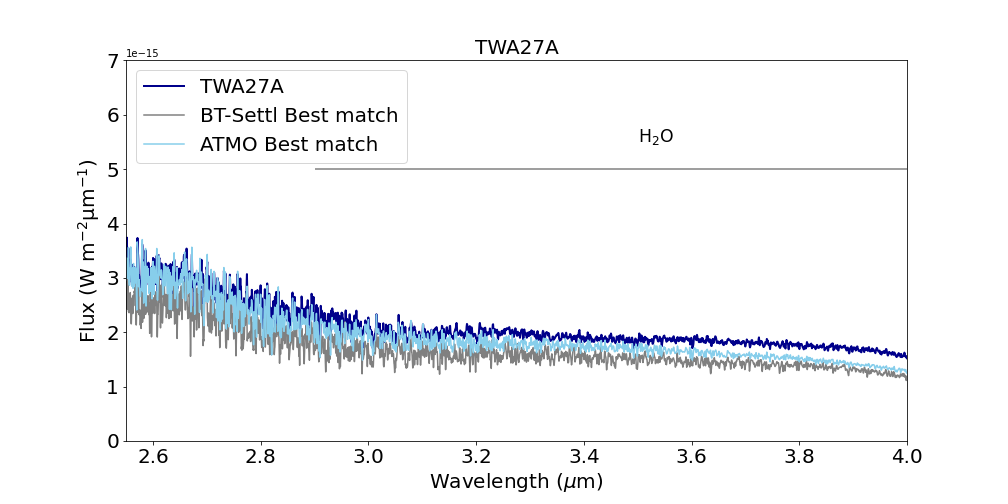}
    \includegraphics[width=0.85\textwidth]{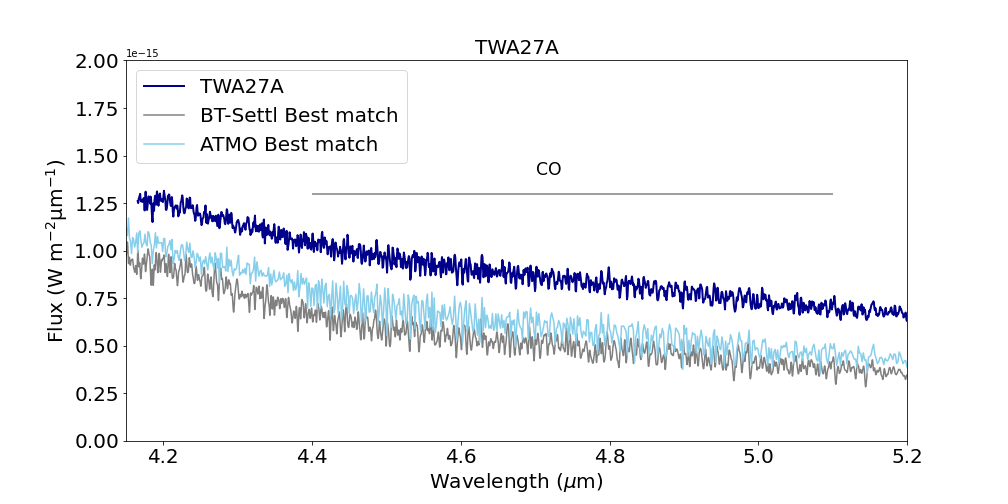}
   \caption{Best fit of the {\tt ATMO} and {\tt BT-Settl CIFIST models} to the TWA~27A 2.0-5.20~$\mu$m spectrum. We highlight the main molecular absorptions present in this wavelength range.}
    \label{TWA27A_BT-Settl+nodisk_ATOM_2.0-5.2}
\end{figure*}

\begin{figure*}
    \centering
    \includegraphics[width=0.85\textwidth]{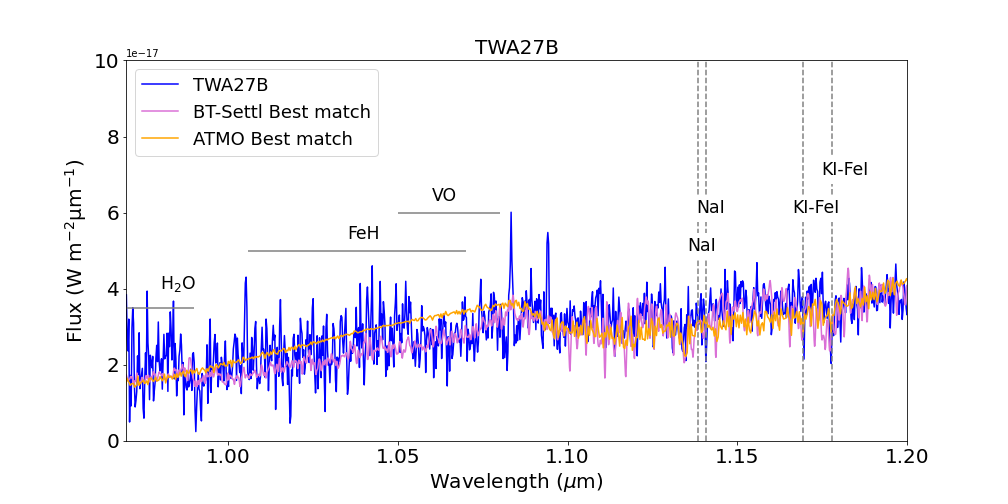}
    \includegraphics[width=0.85\textwidth]{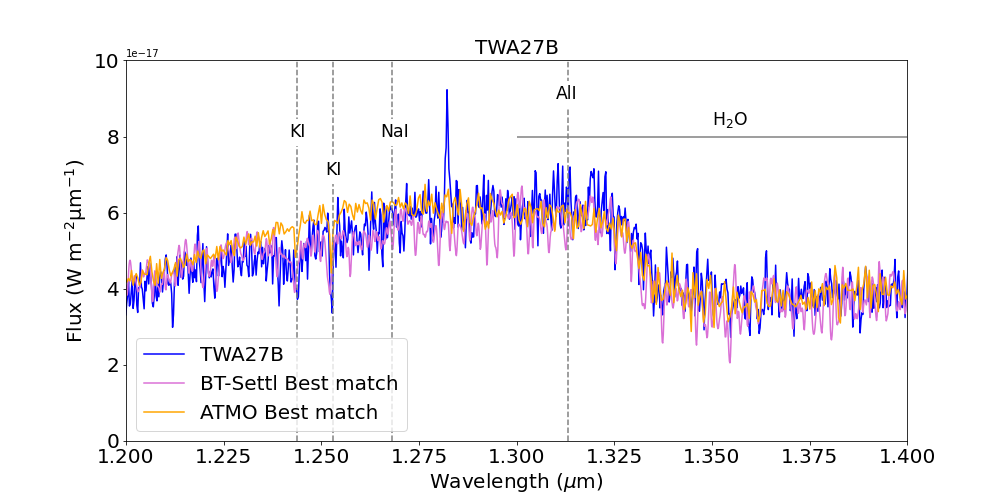}
    \includegraphics[width=0.85\textwidth]{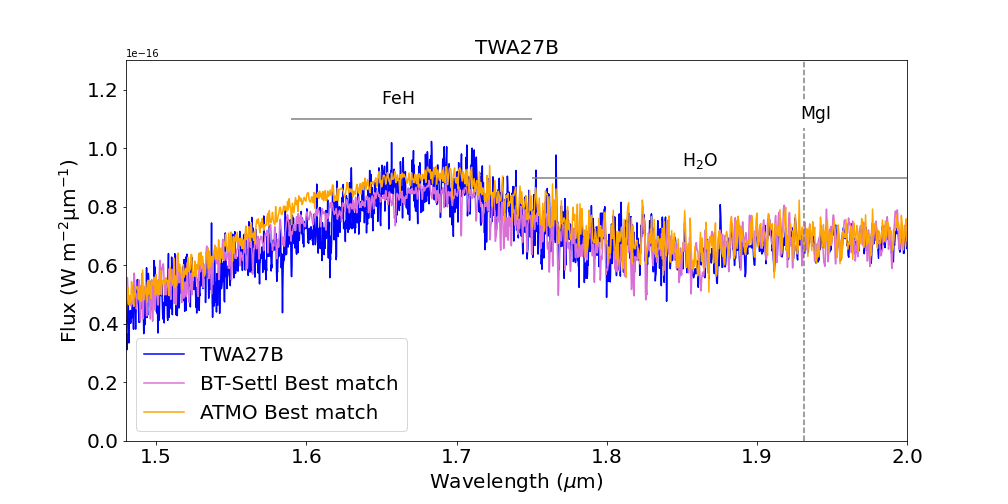}
   \caption{Best fit of the {\tt ATMO} and {\tt BT-Settl models} to the TWA~27B 0.97-2.0~$\mu$m spectrum. We highlight the main atomic lines and molecular absorptions present in this wavelength range.}
    \label{TWA27b_BT-Settl+nodisk_ATOM_0.95-2.0}
\end{figure*}

\begin{figure*}
    \centering
    \includegraphics[width=0.85\textwidth]{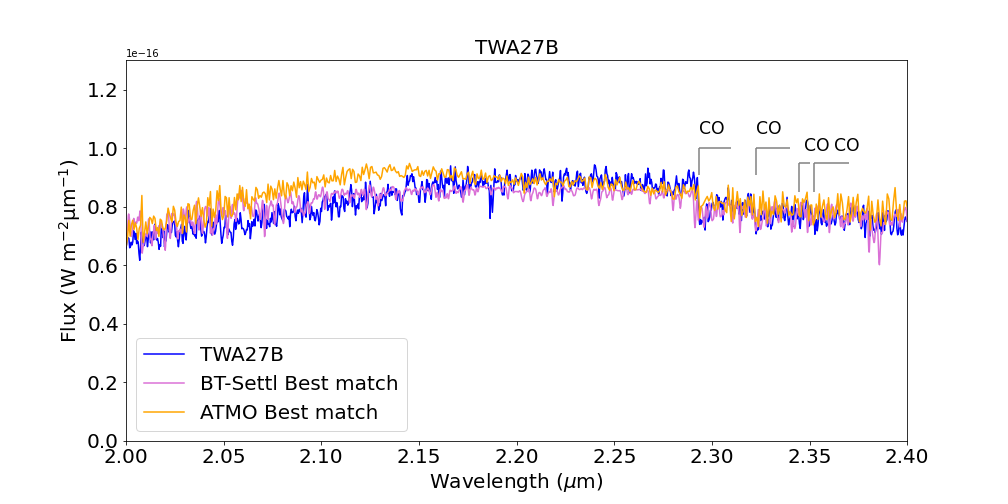}
    \includegraphics[width=0.85\textwidth]{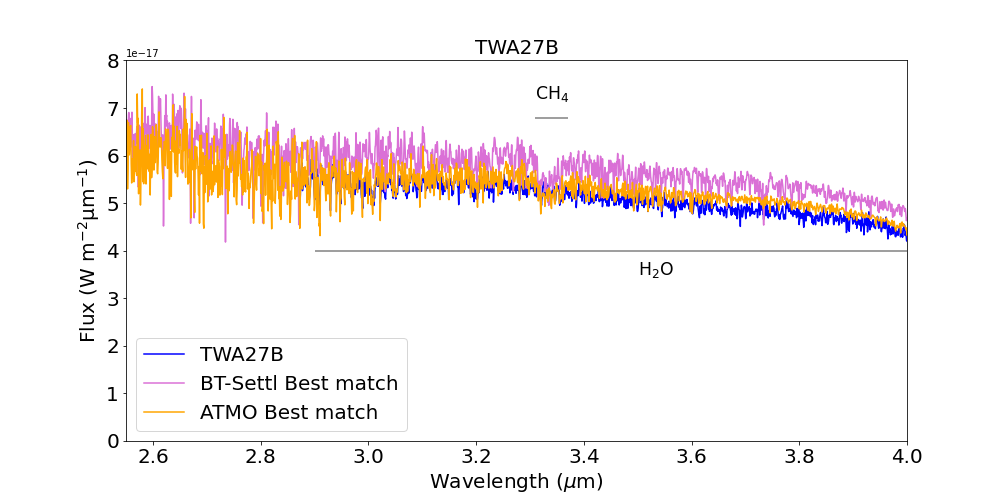}
    \includegraphics[width=0.85\textwidth]{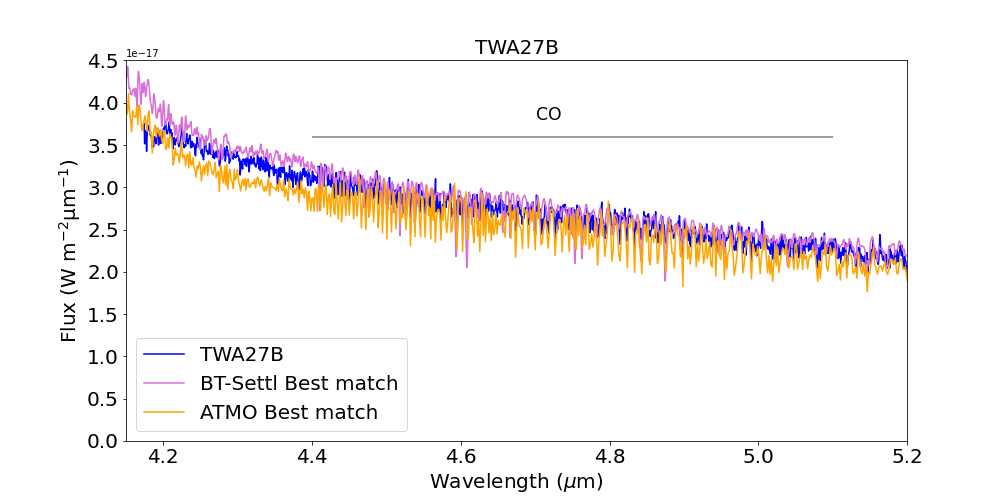}
   \caption{Best fit of the {\tt ATMO} and {\tt BT-Settl models} to the TWA~27B 2.0-5.20~$\mu$m spectrum. We highlight the main molecular absorptions present in this wavelength range.}
    \label{TWA27b_BT-Settl+nodisk_ATOM_2.0-5.2}
\end{figure*}

Comparing how the individual wavelength ranges are reproduced by both models (see Fig. \ref{TWA27A_BT-Settl+nodisk_ATOM_0.95-2.0}
 and \ref{TWA27A_BT-Settl+nodisk_ATOM_2.0-5.2}) we notice that the {\tt BT-Settl CIFIST} and {\tt ATMO} models do not reproduce the depths for the Na\,I and K\,I alkali lines. 
 In the 1.20-1.40~$\mu$m range,  the {\tt ATMO} models over-predict the flux in the water band region between 1.33 and 1.4~$\mu$m. As for TWA~28, the triangular shape of the 1.48-1.68~$\mu$m region is not well-reproduced. The flux between 2.0 and 2.40~$\mu$m is under-estimated by the {\tt BT-Settl CIFIST} model, and beyond the 2.40~$\mu$m, the SED is not well reproduced by the models due to the contribution of the disk as expected. In the 4.0-5.2~$\mu$m region, we notice that the {\tt ATMO} models predict {deeper} CO bands in comparison to the CO bands observed in the NIRSpec spectrum, and the prediction of the {\tt BT-Settl CIFIST} models.


\subsection{TWA~27B Model Comparison}

In the case of TWA~27B JWST/NIRSpec spectrum, we obtained the best match using the {\tt BT-Settl} models to a $\mathrm{T_{eff}}$ = 1318~K, and \textit{log~g} = 3.6 model. The predicted radius for the object is R = 1.15~$\mathrm{R_{Jup}}$, which is likely under-estimated for the young age of TWA~27B. The predicted mass for the object is 1.98~$\mathrm{M_{Jup}}$, which is also under-estimated, if we compare it to other mass estimates from the literature (5$\pm$2~$\mathrm{M_{Jup}}$, e.g. \citealt{Chauvin2005}).

\cite{Luhman2023} compared the JWST/NIRSpec spectrum of TWA~27B to the {\tt ATMO} models finding a $\mathrm{T_{eff}}$=1300~K, \textit{log~g} = 3.5, $\gamma$ = 1.03 (effective adiabatic index), $K_{zz}$ = $\mathrm{10^{5}}$ (diffusion coefficient), and [M/H] = 0.2 (metallicity), with a solar value of C/O. The model predicted a radius of 0.12~$\mathrm{R_{\odot}}$ or 1.16~$\mathrm{R_{Jup}}$, and a mass of 1.6~$\mathrm{M_{Jup}}$, also under-estimated compared with the values reported in the literature. The predicted $\mathrm{T_{eff}}$ and \textit{log~g} are similar to the predicted by the {\tt BT-Settl} models. The best matches are shown in Fig. \ref{best_fit_models-TWA28_TWA27A_TWA27B}.
{A summary of the fundamental physical parameters derived for TWA~27B are summarized in Table \ref{table:summary_results}}.

The spectrum of TWA~27B is noisier than the spectrum of TWA~28 and TWA~27A, thus, some spectral lines are not resolved, nonetheless, we can still compare how well both models reproduce the different molecular bands, and the shapes of the different wavelength ranges. In the range between 0.97 and 1.20~$\mu$m, the {\tt ATMO} model does not reproduce the range between 0.97 and 1.08~$\mu$m, since none of the molecular bands are shown in the modeled spectrum, i.e. the water band at 0.98~$\mu$m, the FeH band at 1.04~$\mu$m and the VO band at 1.06~$\mu$m. 
In the 1.20-1.40~$\mu$m, the {\tt ATMO} model over-predicts the flux between 1.225 and 1.275~$\mu$m. Similarly as for TWA~28 and TWA~27A, {\tt ATMO} does not properly reproduce the triangular shape of the spectrum between 1.50 and 1.7~$\mu$m, and the continuum between 2.00 and 2.15~$\mu$m. From 2.50~$\mu$m, the {\tt BT-Settl} model over-predicts the flux of TWA~27B. In the 2.58-4.00~$\mu$m range, both models predict a $\mathrm{CH_{4}}$ band at 3.35~$\mu$m, which is not observed for TWA~27B, in contrast with other older sources of similar spectral type, like VHS~1256b \cite{Miles2023}, CWISE J050626.96+073842.4, and PSO J318.5338–287--22.860, as mentioned in \cite{Luhman2023}. As for TWA~28 and TWA~27A, in the 4.0-5.2~$\mu$m region, we also notice that the {\tt ATMO} models predict {deeper} CO bands in comparison to the CO bands observed in the NIRSpec spectrum, and the prediction of the {\tt BT-Settl} models.


\begin{table*}
	\small
	\caption{Derived physical parameters from this work for TWA~28, TWA~27A, and TWA~27B {using the {\tt ATMO} and {\tt BT-Settl} models.}}  
	\label{table:summary_results}
	\begin{center}
		\begin{tabular}{llllllllll}
			\hline
			\hline 
				
			Name & $\mathrm{T_{eff}^{a}}$({\tt ATMO})  & $\mathrm{T_{eff}^{a}}$({\tt BT}) & \textit{log~g}$\mathrm{^{b}}$({\tt ATMO}) & \textit{log~g}$\mathrm{^{b}}$({\tt BT}) & R ($\mathrm{R_{Jup}}$) ({\tt ATMO}) &  R ($\mathrm{R_{Jup}}$)({\tt BT}) & M ($\mathrm{M_{Jup}}$) ({\tt ATMO}) &  M ($\mathrm{M_{Jup}}$) ({\tt BT}) \\		
			\hline              
			TWA~28  & {2400} K & {2577} K & {4.0} & {4.0} & {2.90$\pm$0.24} & {2.55$\pm$0.19}  & $\mathrm{32.0^{+67.0}_{-22.2}}$  & $\mathrm{26.0^{+53.5}_{-16.9}}$   \\
            TWA~27A & {2600} K & {2605} K & {4.0} & {4.0} & {2.70$\pm$0.21} & {2.70$\pm$0.20} & $\mathrm{28.1^{+61.8}_{-19.8}} $  & $\mathrm{29.3^{+62.8}_{-19.8}} $    \\
            TWA~27B & {1318} K & {1300} K & {3.5} & {3.6} & {1.16$\pm$0.18}  & {1.15$\pm$0.15}   & $\mathrm{1.6^{+1.4}_{-0.6}}$    & $\mathrm{1.98^{+1.2}_{-0.4}}$      \\
            \hline
				
		\end{tabular}
	\end{center}
 	\begin{tablenotes}
 \small
		\item {a: The uncertainty in $\mathrm{T_{eff}}$ = 100~K for both models. b: The uncertainty in log~g is 0.5 for both models}.\\
    \end{tablenotes}
\end{table*}

\section{Discussion}\label{sec:discussion}

\subsection{Comparison between TWA~28 and TWA~27A}

In this Section, we directly compare the properties of TWA~28 and TWA~27A. {TWA~28 and TWA~27A have similar spectral types and surface gravities}, nevertheless, in the analysis performed in this work, we did find several differences between them. TWA~27A was classified as a M8.0, and TWA~28 was classified as an M9.0 by \cite{Allers2013} using R$\sim$120 IRTF/SpeX spectra.  In contrast, \cite{Venuti2019} classified both of them as M9.0 using VLT/XShooter UVB/VIS/NIR spectra (0.3-2.46~$\mu$m), with a resolution R$\sim$3200-5000. {We adopted the latter spectral classification here since it is based on a {broader} wavelength range spectrum. }

In Section \ref{sec:data_analysis}, we described all the spectral atomic lines and molecular bands found in the spectra of the three objects. For TWA~28 and TWA~27A, we find in general the same wealth of spectral features, nevertheless, when {measuring} the equivalent widths of the most prominent ones, we did notice some differences. {We measured the equivalent width of the K\,I and Na\,I alkali lines for all objects (Section \ref{sec:spectral_indices}), noticing} that the equivalent width of the K\,I lines are not consistent for the K\,I line at 1.177~$\mu$m {(0.27$\pm$0.03~$\mu$m for TWA~28 vs 0.21$\pm$0.03~$\mu$m for TWA~27A)} and at 1.243~$\mu$m {(0.46$\pm$0.02~$\mu$m for TWA~28 vs 0.33$\pm$0.02~$\mu$m for TWA~27A)}, and the equivalent width of the Na\,I line at 1.138~$\mu$m is also not consistent {(0.66$\pm$0.03~$\mu$m for TWA~28 vs 0.55$\pm$0.03~$\mu$m for TWA~27A)} (see Table \ref{table:ew_all_lines_nir}). Similarly, when we compared the gravity scores of both objects (see Table \ref{table:gscores}), we observed that the score that measures the $\mathrm{FeH_{J}}$ is also not consistent {(1.039$\pm$0.001~$\mu$m for TWA~28 vs 1.030$\pm$0.001~$\mu$m for TWA~27A)}. {One of the possible explanations for the differences between} these lines might be veiling, {however} no veiling was reported by \cite{Venuti2019} for TWA~28 and TWA~27A.

Both TWA~28 and TWA~27A are known accretors with an accretion mass of log~$\mathrm{M_{acc}}$ = -12.24 and  log~$\mathrm{M_{acc}}$ = -11.35, respectively \citep{Venuti2019}, but no significant emission lines indicating disk accretion are present in their NIRSpec spectra, {like for example the Pa-$\gamma$ line at 1.094~$\mu$m.}

\begin{figure*}[h]
    \centering
    \includegraphics[width=0.95\textwidth]{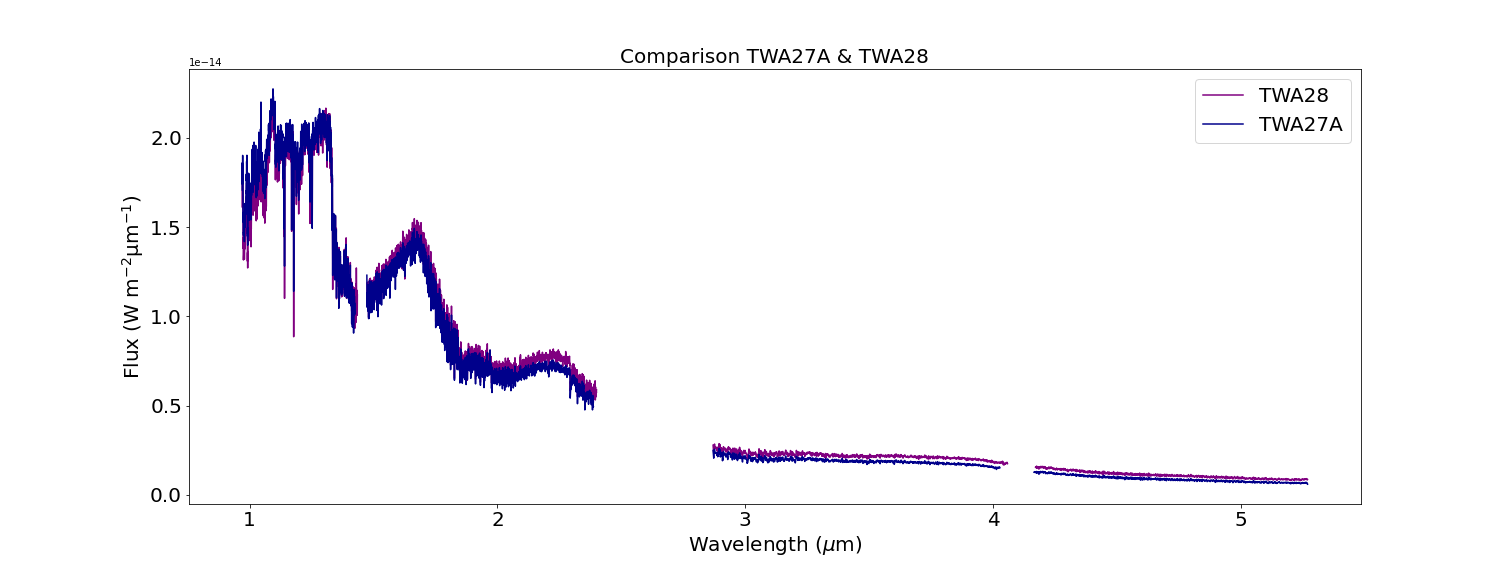}
    \includegraphics[width=0.95\textwidth]{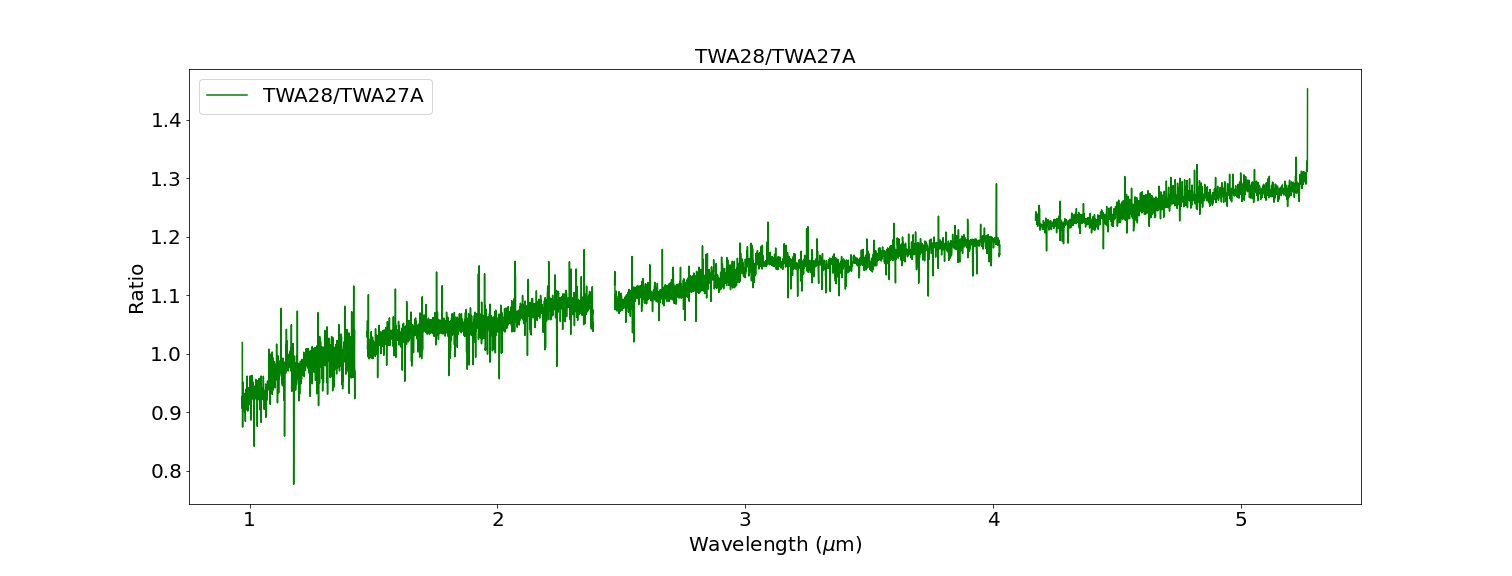}
   \caption{Upper panel: comparison between the 0.97-5.30~$\mu$m NIRSpec spectrum of TWA~28 and TWA~27A. Bottom panel: ratio between the flux of TWA~28 and TWA~27A.}
    \label{TWA28_TWA27A_comparison}
\end{figure*}

\cite{Ricci2017} used ALMA imaging to determine the size of the disk of TWA~27A and TWA~27B, and determined that the disk of TWA~27A is very compact, with a maximum radius of $\sim$10~au, with a mass of 0.1~$\mathrm{M_{\oplus}}$, which is significantly less massive than other brown dwarfs disks ($\mathrm{M_{dust}}$ = 2--6~$\mathrm{M_{\oplus}}$, \citealt{Ricci2012, Testi2016}). \cite{Ricci2017} suggested that the small mass of TWA~27A's disk could be naturally explained by the tidal truncation of the outer disk due to the gravity of TWA~27B. Although \cite{Lodato2005} suggested that TWA~27B might have formed in situ via gravitational fragmentation, both TWA~27A and TWA~27B follow the scaling relation between dust mass and stellar/substellar mass found for more massive brown dwarfs and pre-main sequence stars (i.e. \citealt{Pascucci2016}), suggesting that both objects might have formed from independent protostellar cores. The mass of the disk of TWA~28 and its {extent} has not been reported in the literature. In Fig. \ref{TWA28_TWA27A_comparison} we compare the spectra of both objects, and we provide the ratio of TWA~28 versus TWA~27A. TWA~28 shows more near-infrared excess from 2.00~$\mu$m than TWA~27A, indicating that probably the TWA~28's disk is more massive than TWA~27A's disk. 

{In summary, even though TWA~28 and TWA~27A were expected to have very similar general properties, i.e. similar spectral type, surface gravity, and of course age, there are few differences, namely: the different equivalent widths for the K\,I lines at 1.177~$\mu$m {(0.27$\pm$0.03~$\mu$m for TWA~28 vs 0.21$\pm$0.03~$\mu$m for TWA~27A)} and at 1.243~$\mu$m {(0.46$\pm$0.02~$\mu$m for TWA~28 vs 0.33$\pm$0.02~$\mu$m for TWA~27A)}, and for the Na\,I line at 1.138~$\mu$m {(0.66$\pm$0.03~$\mu$m for TWA~28 vs 0.55$\pm$0.02~$\mu$m for TWA~27A)}, and different  $\mathrm{FeH_{J}}$ gravity score {(1.039$\pm$0.001~$\mu$m for TWA~28 vs 1.030$\pm$0.001~$\mu$m for TWA~27A)}. Slightly different accretion rates as measured by \cite{Venuti2019}, and different infrared excess, probably due to the compact disk found in TWA~27A. \cite{Ricci2017} argues that the compact disk found in TWA~27A is likely due to the tidal truncation of TWA~27B.}

\subsection{Comparison with previous results}

We compared the $\mathrm{T_{eff}}$ we obtained with the {\tt BT-Settl} and {\tt ATMO} models with the results obtained by \cite{Venuti2019} using the ROTFIT code \citep{Frasca2017} for TWA~28 {using optical and near-infrared X-shooter spectra}. {The ROTFIT code simultanteouly finds the spectral type and $v \sin i$ of the target searching into a library of standard spectra}. 

The $\mathrm{T_{eff}}$  reported by \cite{Venuti2019} ($\mathrm{T_{eff}}$ = 2660~K) is closer to {the $\mathrm{T_{eff}}$} reported by the {\tt BT-Settl} models ($\mathrm{T_{eff}}$ = 2577~K), {while the \textit{log~g} {we obtained (\textit{log~g} = 4.1) is consistent with \cite{Venuti2019}}}. \cite{Sgro2021} also derived the $\mathrm{T_{eff}}$ fitting the optical, near-infrared and mid-infrared photometry of TWA~28 with an SED and a disk emission, finding a $\mathrm{T_{eff}}$ = 2360~K, a temperature of the disk's dust of 290~K, and a radius of 0.048~au (or 100~$\mathrm{R_{Jup}}$).  The {\tt BT-Settl CIFIST} models predict a mass for TWA~28 of 26.01$\mathrm{M_{Jup}}$ relatively close with the mass estimated by \cite{Venuti2019} of 20.92$\mathrm{M_{Jup}}$.

Similarly, for TWA~27A, we compared the $\mathrm{T_{eff}}$ we obtained with the {\tt BT-Settl CIFIST} and {\tt ATMO} models with the results obtained by \cite{Venuti2019} using the ROTFIT code \citep{Frasca2017}, and the $\mathrm{T_{eff}}$ they report is consistent with both temperatures derived with {\tt BT-Settl CIFIST} and {\tt ATMO} ($\mathrm{T_{eff}}$ = 2640~K), while the \textit{log~g} they report is slightly smaller than the reported by both models (\textit{log~g} = 3.75). As for TWA~28, \cite{Sgro2021}  derived the $\mathrm{T_{eff}}$ fitting the optical, near-infrared and mid-infrared photometry of TWA~27A with an SED and a disk emission, finding a $\mathrm{T_{eff}}$ = 2550~K, a temperature of the dust of 290~K, and a radius of 0.053~au (or 111~$\mathrm{R_{Jup}}$). For TWA~27A, the {\tt BT-Settl CIFIST} model predicts a mass of 29.3$\mathrm{M_{Jup}}$, higher than the 19.9$\mathrm{M_{Jup}}$ estimated by \cite{Venuti2019}. 

For TWA~27B, \cite{Allers2013} provided a spectral type of L5 using Spex/IRTF spectrum of R$\sim$120, in which no emission lines could be identified, unlike in the NIRSpec spectrum published in \cite{Luhman2023}. Since TWA~27B is faint and only 0."77 from its primary, constraints on the presence of a disk could not be made until \cite{Luhman2023} published the JWST/NIRSpec spectrum. Previous to this publication, using ALMA (Atacama Large Millimeter Array) \cite{Ricci2017} derived a 3$\sigma$ upper limit of $\sim$0.013~$\mathrm{M_{\oplus}}$ ($\sim$1~$\mathrm{M_{Moon}}$) on the mass of dust surrounding TWA 27B. 

If we compare the estimated mass for TWA~27B with the estimation by \cite{Luhman2023} using evolutionary models (0.005~$\mathrm{M_\odot}$ or 5.2~$\mathrm{M_{Jup}}$). The latest is consistent with the mass measured by previous studies \citep{Chauvin2005, Bowler2016}. Thus,  the {\tt BT-Settl CIFIST} provides an underestimation of its mass (1.98$\mathrm{M_{Jup}}$). Finally, using also evolutionary models, \cite{Luhman2023} reported a $\mathrm{T_{eff}}$ of 1200~K which is lower than the temperature derived from both {\tt BT-Settl} and {\tt ATMO} models (1300~K). 


Both {\tt ATMO} and {\tt BT-Settl} models predict the existence of $\mathrm{CH_{4}}$ band between 2.8 and 3.8~$\mu$m \citep{Luhman2023}. For other young L6.0-L7.0 dwarfs, like for CWISE J050626.96+073842.4, PSO J318.5338--22.860 and VHS 1256b, this $\mathrm{CH_{4}}$ band is actually detected, but not in TWA~27B's spectrum. This might be an indication of the L/T transition starting at later spectral types for very young brown dwarfs due to its low surface gravity.  Further observations of other very young brown dwarfs in the 1.0-5.3~$\mu$m wavelength range are needed for confirmation.

Finally, as for TWA~28 and TWA~27A, in the 4.0-5.2~$\mu$m region, we also notice that the {\tt ATMO} models predict  {deeper} CO bands in comparison to the CO bands observed in the NIRSpec spectrum, and the prediction of the {\tt BT-Settl} models. \cite{Luhman2023} already observed the weakness of the CO band at 4.8~$\mu$m in the TWA~27B spectrum with respect to the {\tt ATMO} model. \cite{Luhman2023} discussed that the inclusion of clouds in the model might weaken the $\mathrm{CH_{4}}$ band at 3.35~$\mu$m and the CO band at 4.8~$\mu$m. {However, the {\tt BT-Settl} cloudy models still predict the existence of the $\mathrm{CH_{4}}$ band at 3.35~$\mu$m}. 
{The lack of $\mathrm{CH_{4}}$ bands in the spectrum of TWA~27B might suggest that the L/T transition of very young ($<$10~Myr) brown dwarfs start at later spectral types than for older brown dwarfs. 
Further observations of other very young brown dwarfs in the 1.0-5.3~$\mu$m wavelength range are needed for confirmation. }

TWA~27B shows photometric variability with HST/WFC3 \citep{Zhou2016} by with an amplitude detected a variability amplitude in the 1.36\% in the F125W and 0.78\% in the F160W filters, respectively, with a period of 10.7~hr, suggesting the existence of heterogeneous clouds in the atmosphere of TWA~27B. No rotational modulations have been reported for TWA~28 and TWA~27A potentially indicating the existence of clouds, but since both show signs of magnetic activity due to the presence of H-$\alpha$ in their optical spectra \citep{Venuti2019}, it would be challenging to disentangle if the cause of the rotational modulations would be due to clouds or magnetic activity.


\section{Conclusions}\label{sec:conclusions}

We present the 0.97--5.27~$\mu$m JWST/NIRSpec medium-resolution spectra of TWA~28, TWA~27A and TWA~27B obtained as a part of the guaranteed time observation program 1270 (PI: S. Birkmann). In the following we summarize our main results: 

\begin{enumerate}

    \item We identified the majority of the atomic lines present in the 0.97--5.27~$\mu$m spectra for TWA~28, TWA~27A and TWA~27B. The most prominent atomic lines detected are the K\,I  alkali lines at 1.169, 1.177, 1.243, and 1.253~$\mu$m, and the Na\,I line at 1.138~$\mu$m. We detected less prominent lines like the Fe\,I at 1.180~$\mu$m, the Al\,I at 1.313~$\mu$m, and the Mg\,I at 1.931~$\mu$m, among others. 

    \item We identified the majority of the molecular bands present in the 0.97--5.27~$\mu$m spectra for TWA~28, TWA~27A and TWA~27B. The most prominent bands are water bands at 1.40, 1.85, and 2.80~$\mu$m, the CO bands between 1.29 and 2.40~$\mu$m, and the CO band at 4.8~$\mu$m.

    \item We also detected emission lines indicative of accretion processes in TWA~27B (Paschen $\beta$, $\gamma$, $\alpha$), and He\,I in TWA~27B. \cite{Luhman2023} and \cite{Marleau2024} derived an accretion rate of $\dot{M}$~=~$10^{-13}-10^{-12}M_{\odot}\, yr^{-1}$, which correlates with the expected  accretion rate extrapolating the mass of TWA~27B. {The weak accretion rate implies that formation is likely over \citep{Marleau2024}}. TWA~28 and TWA~27A do not show significant emission lines in their NIRSpec spectra, but they are known accretors \cite{Venuti2019}. TWA~28 and TWA~27A show near-infrared excess from $\sim$2.5~$\mu$m on, supporting the existence of a cicumstellar disk around them as reported by \cite{Venuti2019}.
    
     \item We measured the gravity scores as presented in \cite{Allers2013} using the K\,I and Na\,I alkali lines, and some bands, like the FeH, the $H$-band and the VO band. We concluded that all scores point to all three objects to have very low surface gravity, as expected for $\sim$10~Myr objects, such TWA~28, TWA~27A and TWA~27B.

     \item We compared the three objects to the cloudless {\tt ATMO} and cloudy {\tt BT-Settl} or {\tt BT-Settl CIFIST} atmospheric models. We concluded that both models provide similar effective temperatures and surface gravities for TWA~27A and TWA~27B. For TWA~28, the {\tt BT-Settl CIFIST} models find a $\sim$200~K warmer effective temperature than {\tt ATMO}.

     \item The {\tt BT-Settl CIFIST} models predict a mass for TWA~28 of 26.0~$\mathrm{M_{Jup}}$ close the mass estimated by \cite{Venuti2019} of 20.9~$\mathrm{M_{Jup}}$. For TWA~27A, the {\tt BT-Settl CIFIST} model predicts a mass of 29.3~$\mathrm{M_{Jup}}$, higher than the 19.9~$\mathrm{M_{Jup}}$ estimated by \cite{Venuti2019}. For TWA~27B, the {\tt BT-Settl} models under-predict its mass (1.98~$\mathrm{M_{Jup}}$), which has been estimated to be $\mathrm{5\pm2}$~$\mathrm{M_{Jup}}$ by previous works \citep{Luhman2023, Chauvin2005, Bowler2016}.

     \item The {\tt ATMO} models overpredict the CO bands at 4.8~$\mu$m for all three objects. Adding clouds to the model might decrease the depth of these bands, and might provide a more accurate estimation of those. The cloudly {\tt BT-Settl} models provide a more accurate prediction of the depth of the band.

    \item For TWA~27B, both models predict the existence of $\mathrm{CH_{4}}$ bands at 2.15~$\mu$m and at 3.35~$\mu$m. Those are not present in the TWA~27B NIRSpec spectrum, but they were found in older, similar spectral type brown dwarfs, like PSO~318 and VHS~1256b \citep{Luhman2023}. The lack of $\mathrm{CH_{4}}$ bands in the spectrum of TWA~27B might suggest that the L/T transition of very young ($<$10~Myr) brown dwarfs start at later spectral types than for older brown dwarfs. Further observations of other very young brown dwarfs in the 1.0-5.3~$\mu$m wavelength range are needed for confirmation. 
    
\end{enumerate}

\begin{acknowledgments}

Based on observations collected with the \textit{NIRSpec} Instrument onbard \textit{James Webb Space Telescope} under the GTO program 1270 (P.I. S. Birkmann). 

\end{acknowledgments}

%

\vspace{5mm}
\facilities{JWST (NIRSpec)}


\software{astropy \citep{2013A&A...558A..33A,2018AJ....156..123A, 2022ApJ...935..167A}}



\appendix

In this appendix we show the best fitting parameters obtained using the package {\tt species} to the {\tt BT-Settl} atmospheric models.

\begin{figure*}[h]
    \centering
    \includegraphics[width=0.95\textwidth]{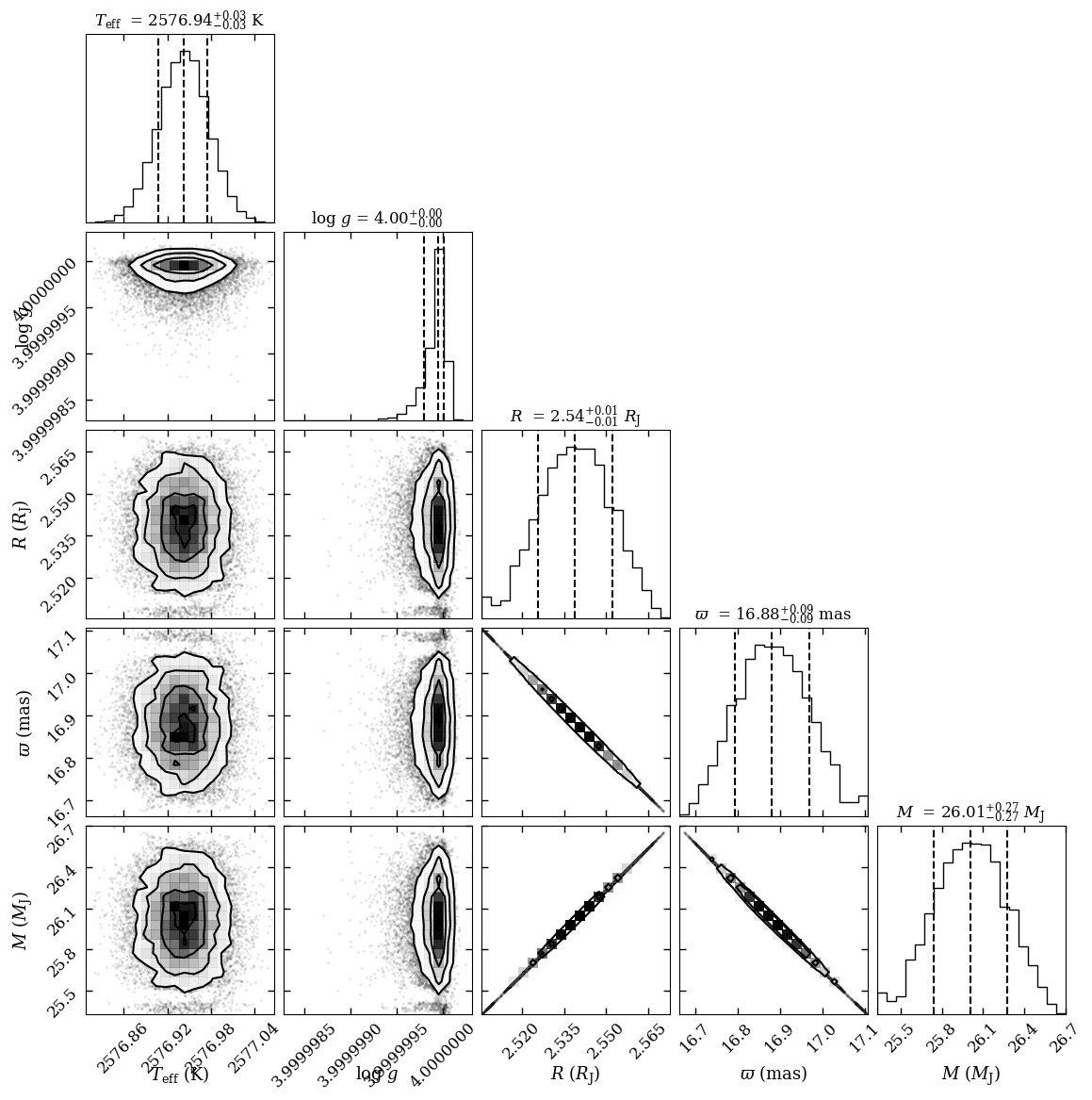}
   \caption{{MCMC corner plot with the best fit parameters (effective temperature, $\mathrm{T_{eff}}$; surface gravity, log~g; radius, R; parallax, $\omega$; and mass, M) derived with the best fit using the {\tt BT-Settl CIFIST} models for TWA~28. The vertical lines show the 1-$\sigma$ uncertainties of each parameter as derived by the MCMC fitting.}}
    \label{multinest-TWA28}
\end{figure*}

\begin{figure*}[h]
    \centering
    \includegraphics[width=0.95\textwidth]{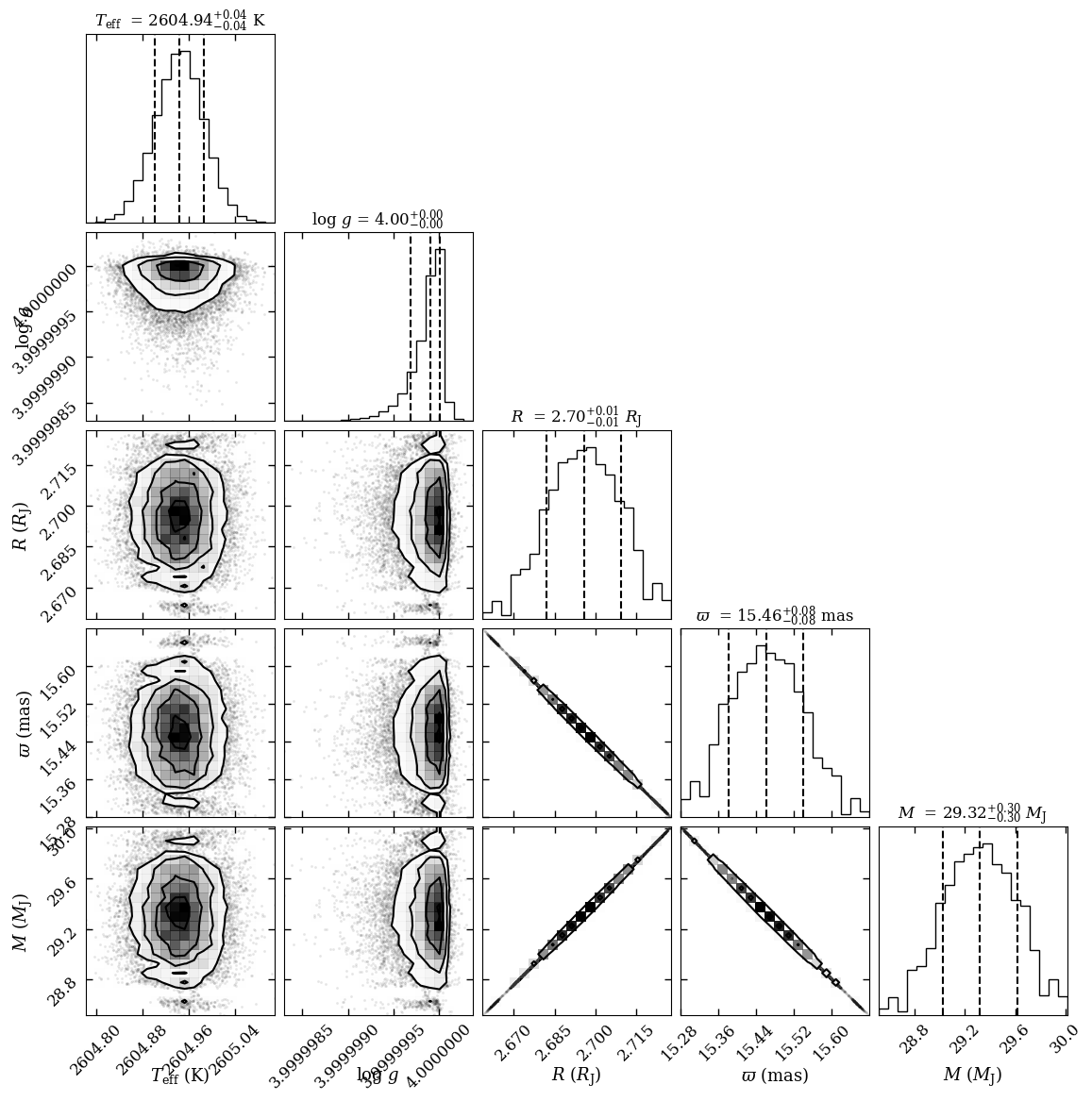}
   \caption{{MCMC corner plot with the best fit parameters (effective temperature, $\mathrm{T_{eff}}$; surface gravity, log~g; radius, R; parallax, $\omega$; and mass, M) derived with the best fit using the {\tt BT-Settl CIFIST} models for TWA~27A. The vertical lines show the 1-$\sigma$ uncertainties of each parameter as derived by the MCMC fitting.}}
    \label{multinest-TWA27A}
\end{figure*}

\begin{figure*}[h]
    \centering
    \includegraphics[width=0.95\textwidth]{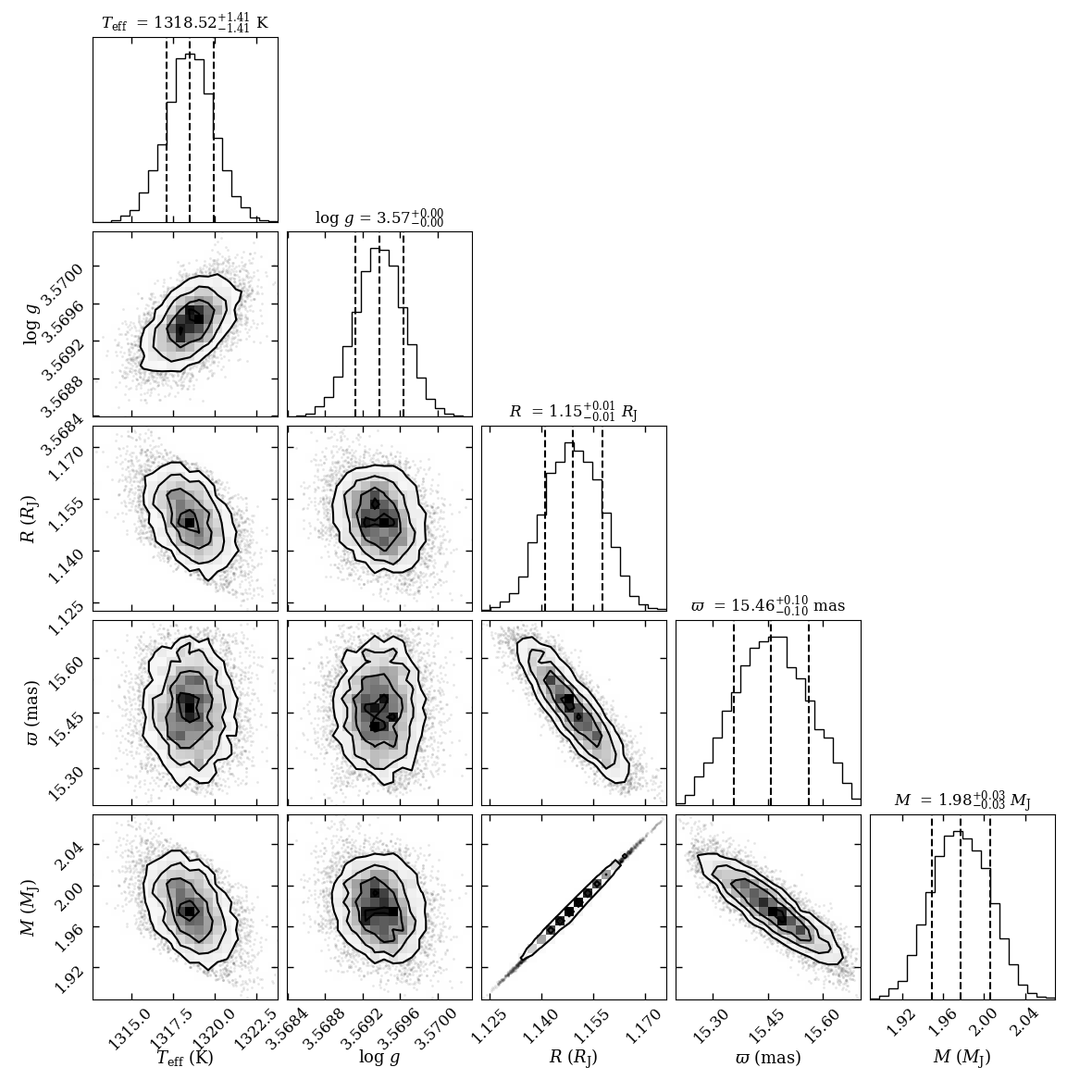}
   \caption{{MCMC corner plot with the best fit parameters (effective temperature, $\mathrm{T_{eff}}$; surface gravity, log~g; radius, R; parallax, $\omega$; and mass, M) derived with the best fit using the {\tt BT-Settl} models for TWA~27B. The vertical lines show the 1-$\sigma$ uncertainties of each parameter as derived by the MCMC fitting.}}
    \label{multinest-TWA27b}
\end{figure*}


\bibliography{TWA27Ab_28}{}
\bibliographystyle{aasjournal}



\end{document}